\documentclass[journal=jctcce,manuscript=article]{achemso}

\setkeys{acs}{keywords = true}

\usepackage[version=3]{mhchem} 

\usepackage{framed} 
\newenvironment{highlights}{
  \begin{framed}
  \noindent\textbf{Highlights:}\vspace{0.5em}
  \begin{itemize}
}{
  \end{itemize}
  \end{framed}
}

\usepackage{subcaption}

\usepackage{booktabs}   
\usepackage{multirow}   

\usepackage{float}  

\author{Oscar A. Oviedo}

\email{o.a.oviedo@unc.edu.ar}
\affiliation { Consejo Nacional de Investigaciones Científicas y Técnicas (CONICET), Instituto de Investigaciones en Fisicoquímica de Córdoba (INFIQC), Universidad Nacional de Córdoba, Facultad de Ciencias Químicas, Departamento de Química Teórica y Computacional, X5000HUA, Córdoba, Argentina.}

\title[]
{Short-Term Regional Electricity Demand Forecasting in Argentina Using LSTM Networks}


\begin{document}

\begin{tocentry}
  \begin{center}
     \includegraphics[width=\textwidth]{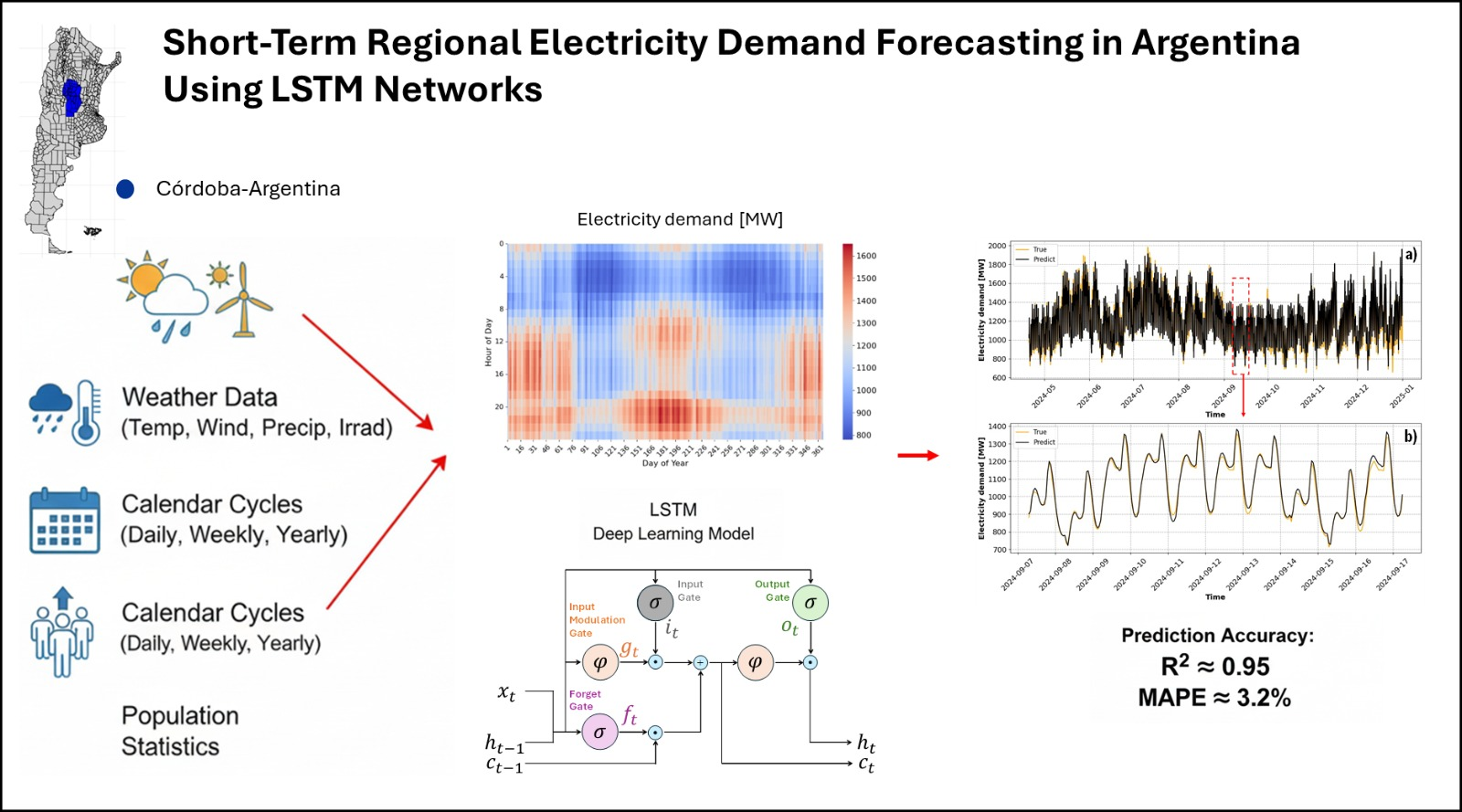}
  \end{center}
\end{tocentry}

\begin{abstract}
This study presents the development and optimization of a deep learning model based on Long Short-Term Memory (LSTM) networks to predict short-term hourly electricity demand in Córdoba, Argentina. Integrating historical consumption data with exogenous variables (climatic factors, temporal cycles, and demographic statistics), the model achieved high predictive precision, with a mean absolute percentage error of 3.20\% and a determination coefficient of 0.95. The inclusion of periodic temporal encodings and weather variables proved crucial to capture seasonal patterns and extreme consumption events, enhancing the robustness and generalizability of the model. In addition to the design and hyperparameter optimization of the LSTM architecture, two complementary analyses were carried out: (i) an interpretability study using Random Forest regression to quantify the relative importance of exogenous drivers, and (ii) an evaluation of model performance in predicting the timing of daily demand maxima and minima, achieving exact-hour accuracy in more than two-thirds of the test days and within ±1 hour in over 90\% of cases. Together, these results highlight both the predictive accuracy and operational relevance of the proposed framework, providing valuable insights for grid operators seeking optimized planning and control strategies under diverse demand scenarios.
\end{abstract}

\begin{highlights}
\item Developed a high-precision LSTM model for short-term hourly electricity demand forecasting.
\item Achieved state-of-the-art accuracy with MAPE = 3.20\% and $R^2=0.95$.
\item Incorporated weather variables, temporal encodings, and demographic factors as exogenous inputs.
\item Demonstrated robustness in capturing both seasonal patterns and extreme demand events.
\item Provided open-access datasets and baseline models to ensure  transparency and reproducibility.
\end{highlights}

\section{Introduction}
Power systems are large-scale infrastructures that integrate generators, transmission networks, and distribution systems to deliver electricity continuously, efficiently, and reliably to end users \cite{kundur1994, Hong2020} . Their stability depends on maintaining a real-time balance between generation and demand \cite{Khaloie2025, DeFelice2013} , as large-scale electricity storage remains prohibitively expensive \cite{Mystakidis2024} . Insufficient electricity generation may result in blackouts, economic disruption, and compromised provision of essential services, including healthcare, education, and communication. In contrast, excessive generation exacerbates environmental impacts, destabilizes energy markets by reducing revenues and discouraging investment, and places additional stress on transmission and distribution infrastructure. Accurate short-term forecasting, typically a few hours ahead, is therefore essential for system operators to anticipate demand fluctuations and schedule generation dispatch. Beyond preventing contingencies and unnecessary costs, reliable forecasts also enhance the technical and economic performance of power systems, where electricity stands as the most volatile commodity \cite{daSilvaLeite2023, Frommel2014} .

The economic impact of accurate energy forecasting, along with its strategic role in modern power system planning, has fueled a surge of scientific interest over the past decade, reflected in an exponentially growing body of academic research \cite{Yang2021} . This trend has even inspired international competitions focused on methodological advances, such as the Global Energy Forecasting Competition (GEFCom), which accelerated the development of advanced electricity demand forecasting techniques \cite{HongFan2016} . Since its last edition in 2017, GEFCom has left a legacy, with its influence extending to derivative events such as the BigDEAL Forecasting Competitions \cite{ShuklaHong2024} . Yet, participation remains geographically uneven: in the 2022 BigDEAL Challenge, for instance, 121 contestants from 27 countries took part, but none from Latin America. This absence contrasts with the region’s rapidly growing energy demand, with both total consumption and electricity use projected to rise significantly over the next decade \cite{IEA2025} . At the same time, Latin America is undergoing a structural transformation of its energy mix, marked by a strong shift toward renewables, creating both a pressing need and a unique opportunity to strengthen local capabilities in energy modeling, forecasting, and planning.

To illustrate the local economic impact, consider Argentine on Augst 25, 2025, when moderate cold drove electricity demand to 17,061 MW (at 8:00 am), compared with a forecast of 17,345 MW. This 284 MW gap (approx. 1.64\%), sustained over the day, represents an unanticipated deviation of 6,816 MWh/day. At a stational average wholesale price of 54 USD/MWh (from CAMMESA, June 2025), the forecast error translates into a potential economic impact of about 368,064 USD/day. Such deviations may require unplanned generation or spot market purchases, undermining both the technical and economic efficiency of the national power system \cite{CAMMESA2025} .

Beyond the economic impacts of forecast errors, it is essential to understand the external drivers of electricity demand, particularly those related to weather conditions and exceptional events. Weather conditions have a significant influence on the temporal evolution of electricity demand \cite{Hong2020, Mystakidis2024, Yang2021, HongFan2016, ShuklaHong2024} . Variables such as ambient temperature, relative humidity, wind speed (and direction), and cloud cover directly affect consumers’ energy usage patterns, particularly in residential and commercial sectors. Among these climatic factors, temperature is identified as the primary determinant due to its direct impact on heating and cooling systems \cite{Staffell2023, Apadula2012, Sailor1997} . Secondary variables, including heating and cooling degree-days, have also been incorporated to assess the impact of climatic conditions on the accuracy of demand prediction models \cite{Sailor2001, PagaGurer2018} . Additionally, extraordinary events such as national or local holidays, religious celebrations (e.g., Easter, Christmas), or specific operational conditions in the energy market (scheduled start-up or shutdown of large industrial consumers for maintenance, vacations, or supply restrictions), can generate significant deviations from typical demand patterns \cite{Ziel2018, Lopez2019, Lopez2022} . The interaction between weather conditions and extraordinary events introduces non-linear dynamics in electricity demand. For example, the same temperature may lead to different consumption levels if it coincides with a holiday, a prolonged heat wave, or an industrial activity disruption. This combination of exogenous factors, whose relationships are not always additive or independent, generates synergistic or modulating effects that are difficult to capture with traditional linear models. Beyond demand deviations, extreme weather events, like storms, heat waves, freezing conditions impact broader energy system components often causing blackouts or widespread disturbances \cite{Goncalves2024} .

Forecasting electricity demand in Argentina presents additional challenges due to the country’s vast geographic, climatic, and socio-economic diversity. Its extensive territory results in significant climate variations, directly affecting overall energy consumption patterns (electricity, natural gas, and other fuels). These factors are compounded by differences in population density, urbanization levels, access to electricity (and other energy vectors), and regional variations in productive activity. Such heterogeneities make “typical” demand behavior neither uniform nor predictable with simple centralized models, requiring approaches that integrate local information, regional weather data, and specific hourly consumption profiles. Developing forecasting models that capture these particularities is essential for efficient planning, especially under high-demand or vulnerable conditions. In the literature, approaches are categorized into national and regional levels, and can be constructed bottom-up, top-down, middle-out, or through forecast reconciliation \cite{SilveiraNetto2023,Athanasopoulos2024} .

A wide variety of forescast models have been proposed, grouped in literature mainly in three categories of energy consumption study, according to the time horizon of their forecast. Long-term forecasts (5-20 years) mainly applied to resource management and development investments. Medium-term forecasts (one month to 5 years) is typically used for planning resources and rates for energy production, and short-term forecast (one hour to one week) is mostly used for analysis and scheduling of the distribution network \cite{Ghalehkhondabi2017} . Regarding analysis methodology, methods can be classified in two categories based on the characteristics of the data: causal and historical \cite{AlAlawi1996} . In causal methods, the cause-effect relationship between energy consumption, as an output, and some input variables, such as economic, social and climatic factors are considered. Regression models \cite{Mukhopadhyay2017,Suganthi2012} and artificial neural networks (ANN) \cite{Imtiaz2006, Ahmed2012, Parkpoom2008, Ghiassi2008, Kavaklioglu2011, Hassan2015, AlHamadi2006, Lam1998, AlQahtani2013} are the most frequently used methods to forecast energy demand. On the other hand, methods based on historical data use the past values of a variable to forecast its future values. Time series \cite{Simmhan2013, GarciaAscanio2010, Wang2011} , Gray prediction \cite{Rui2013, Walter2014, Akay2007, Zhou2006} and autoregressive models \cite{Zhaozheng2010} are among these methods.

The development of artificial intelligence (AI) and machine learning (ML) techniques has undoubtedly advanced energy forecasting \cite{Hong2020, Mystakidis2024} . ML-based models, such as recurrent neural networks (RNNs), have emerged as promising alternatives to reduce estimation errors and optimize power system operations \cite{Khaloie2025} . While RNNs have attracted the most attention, they are not the only approaches. Data-driven methods, multivariate statistical analysis, and natural language processing are also being explored to improve prediction quality in highly dynamic and variable electricity systems \cite{Staffell2023} . Classical time series models, such as ARIMA (AutoRegressive Integrated Moving Average) and its seasonal extension SARIMA, have been widely applied in short-term forecasting, particularly when stationary series with well-defined seasonality are available \cite{Hong2020} . These approaches offer interpretability and low computational cost, although their ability to capture non-linear relationships and multiple exogenous variables is limited compared to NN-based models, as discussed in Tian et al. \cite{Tian2018} .

Qureshi et al. \cite{Qureshi2024} proposed an optimized LSTM model for predicting hourly electricity demand in a U.S. hospital, reporting a coefficient of determination of approximately 0.95. However, the lack of detail regarding the model’s architecture and hyperparameter settings makes the results difficult to reproduce. In a related study, Caicedo-Vivas and Alfonso-Morales \cite{CaicedoVivas2023} presented an LSTM-based neural network for short-term load forecasting applied to a Colombian grid operator. Their model incorporated historical load data along with calendar features such as holidays and the current month. The LSTM layers were used to process multiple load measurements simultaneously, while the additional features were concatenated with the LSTM outputs before being passed to a fully connected layer to produce the final forecasts. Despite the presence of atypical demand patterns linked to social events, their model demonstrated robust predictive performance, with a best-week MAPE of 1.65\% and a worst-week MAPE of 26.22\%.

Extending beyond pure LSTM models, hybrid architectures that combine convolutional and recurrent layers have also been investigated. Tian et al. \cite{Tian2018} compared CNNs, LSTMs, and a hybrid CNN–LSTM model for load forecasting in Northern Italy. Their findings revealed that the hybrid approach consistently outperformed the standalone models, achieving a MAPE of 3.96\% compared to 4.80\% for LSTM and 4.72\% for CNN. Similarly, Teleron et al. \cite{Teleron2025} assessed the performance of Bidirectional-LSTM and CNN–LSTM models using a public dataset from transmission system operators. The CNN–LSTM achieved superior predictive accuracy, with $R^2=0.97$ versus $R^2=0.94$ for BiLSTM, and MAPEs of 1.80\% and 2.52\%, respectively. Although these results are encouraging, the limited reporting of hyperparameter configurations and optimization strategies reduces the transparency and reproducibility of the studies. Demir and Gunal \cite{Demir2025} evaluated LSTM, CNN, and CNN–LSTM models, reporting a MAPE of 4\%. Along the same lines, Faisal et al. \cite{Faisal2022} compared similar architectures and observed comparable performance; however, when incorporating an additional communication channel (C3), their model achieved an exceptionally low MAPE of 0.456\%. This result, while striking, should be interpreted with caution since the dataset used was highly curated, which may have artificially improved the model’s accuracy.

Few studies have addressed the problem of energy demand forecasting at the national level in Argentina \cite{Malvicino2015, Gutierrez2013, Abril2000, Ramos2019}. Mastronardi et al. in 2016 \cite{Mastronardi2016} shown a comparative analysis of regions using econometric models to estimate the influence of temperature on the demand for electrical energy, reaching an $R^2=0.84$. In the case of the Argentine power system, Wholesale Electricity Market Administrator Company (called CAMMESA in Argentina) has implemented hourly demand estimation routines that are essential for the reliable operation of the interconnected grid \cite{CAMMESAWebsite} . Likewise, the National Gas Regulatory Entity \cite{ENARGASWebsite} (known as ENARGAS in Argentina) has advanced the development of natural gas demand forecasting models, combining classical statistical techniques with ML-approaches, thus providing valuable methodological precedents applicable to the energy sector \cite{ENARGAS2020} . Previous studies in Argentina, such as that of Uhrig et al. \cite{Uhrig2025} applied to electricity demand in the province of Entre Ríos, have demonstrated that LSTM-based models can capture nonlinear dependencies achieving strong predictive performance ($R^2=0.77$) without prior feature engineering.

The present study proposes a short-term electricity demand forecasting model for the province of Córdoba-Argentina, developed using deep learning techniques based on recurrent neural networks with LSTM architecture. The model was trained with historical consumption records, climate variables, and demographic statistics, carefully integrated from public data sources without preprocessing. Emphasis is placed on the model’s ability to anticipate high-demand peaks, which represent critical moments of both operational and economic impact.

\section{Data Processing}
The following section describes the origin, extraction, processing, and characteristics of the data used, along with the model employed for forecasting.

\subsection{Population and distribution}
The province of Córdoba is one of the 23 provinces that make up the Argentine. As of December 2024, its population is estimated at 3.925 million inhabitants, representing 8.3\% of the total population of Argentina. Approximately 37.8\% of the population is concentrated in the city of Córdoba, whose metropolitan area constitutes the second-largest urban agglomeration in the country, after Greater Buenos Aires. Population data were obtained from INDEC \cite{INDEC2025} , by the data resulting from the National Population, Household and Housing Census 2010. Figure 1 shows a demographic map of Argentina and the Province of Córdoba, together with a color palette indicating the population density of the 26 departments.

\begin{figure}[htb!]
     \centering
     \includegraphics[width=0.9\columnwidth, keepaspectratio=true]{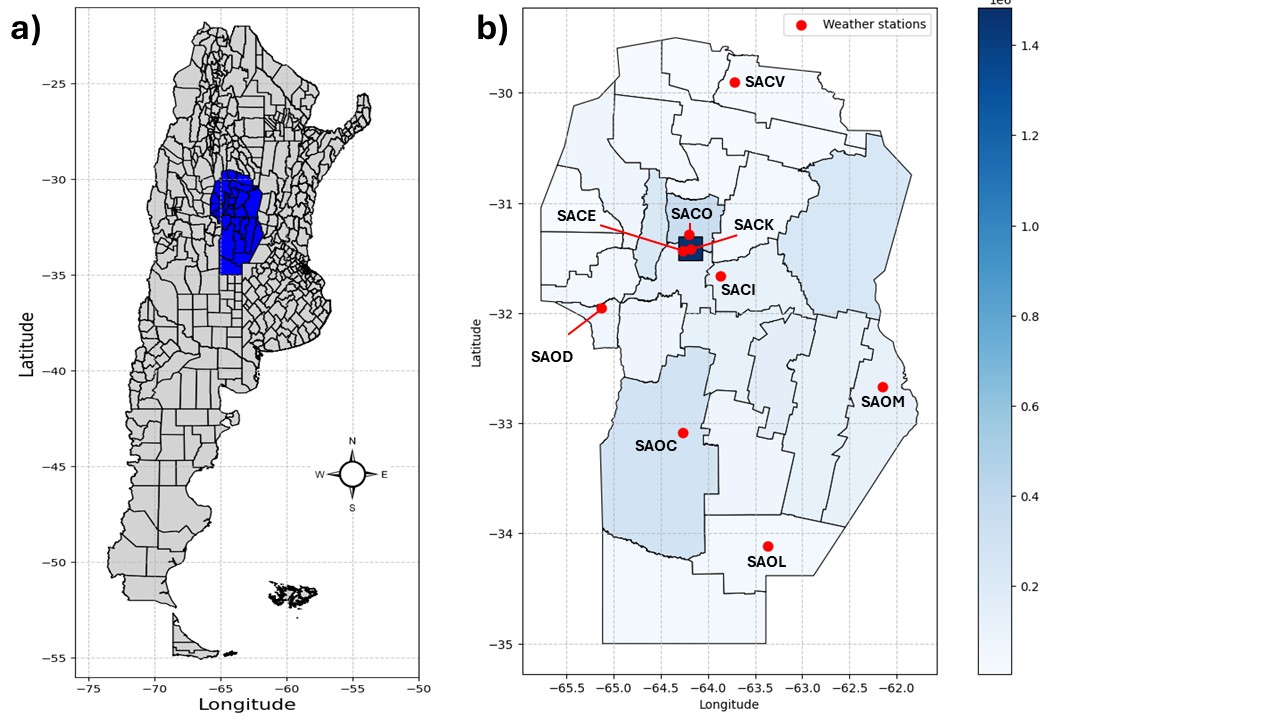}
     \caption{ a) Location of Córdoba (blue) within Argentina. b) The Mediterranean province of Córdoba showing population by department for the year 2024. Weather stations are indicated with red dots.
     }
\label{fig:Fig1}
\end{figure}

\subsection{Meteorological and Climate Variables}
\begin{figure}[htb!]
    \centering
    \begin{subfigure}[b]{0.75\textwidth}
        \centering
        \includegraphics[width=\textwidth]{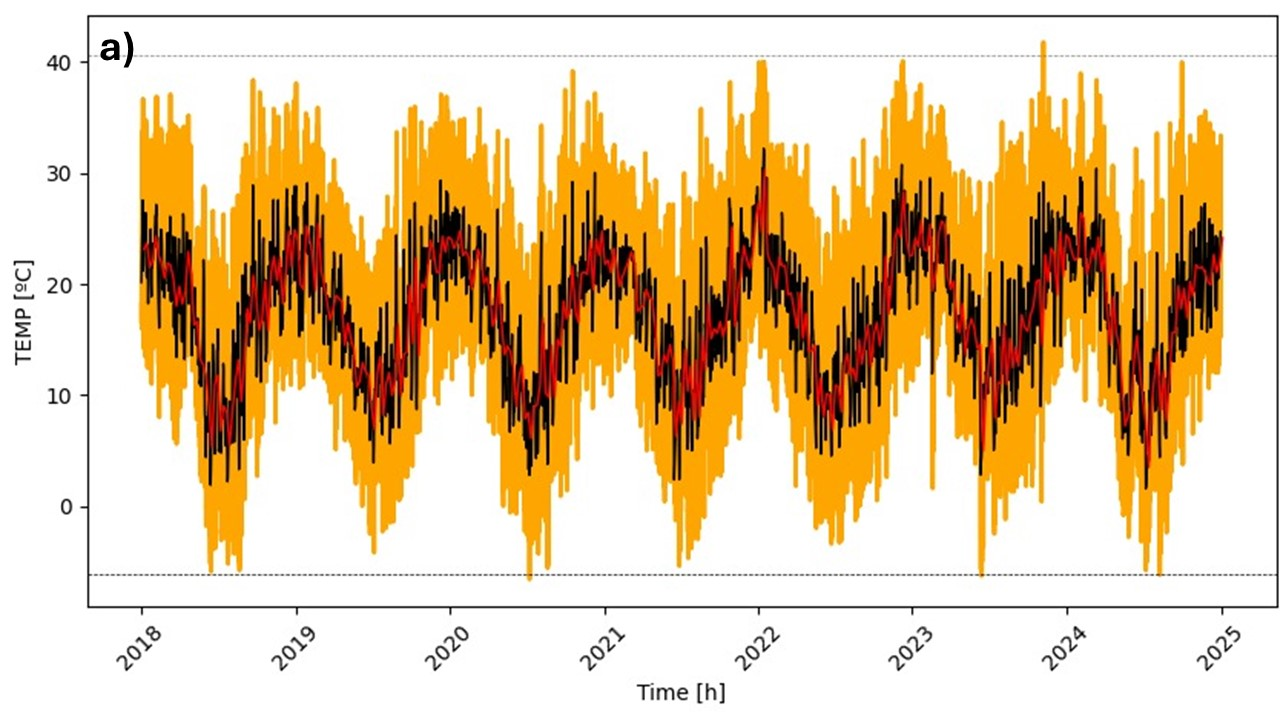}
        \label{fig:Fig2temp}
    \end{subfigure}

    \vspace{0.5em} 

    \begin{subfigure}[b]{0.75\textwidth}
        \centering
        \includegraphics[width=\textwidth]{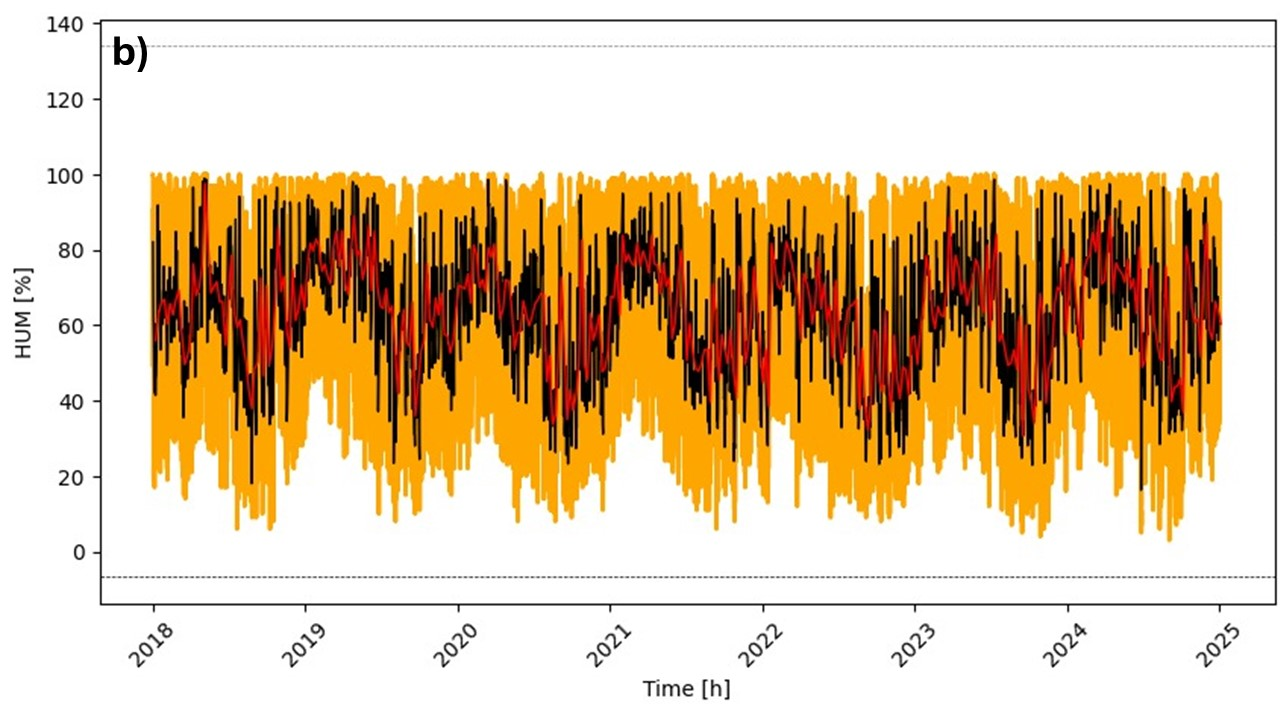}
        \label{fig:Fig2hum}
    \end{subfigure}

    \caption{ a) Temperature [°C] and b) relative humidity [\%], both unprocessed, showing hourly measurements, daily and weekly averages obtained from the SMN (SACO). Horizontal dotted lines indicate the upper and lower bounds (3*$\sigma$), used to identify outliers.}
    \label{fig:Fig2}
\end{figure}

The National Meteorological Service (called SMN in Argentina) \cite{SMNWebsite} provides, through its “open data” section, information from nine meteorological stations in the province of Córdoba, shown in Figure \ref{fig:Fig1}b). This data is available at an hourly resolution and has been recorded from January 2018 to December 2024. From this point onward, the data from these stations will be referred to simply as MD (Meteorological Data), including temperature (TEMP), relative humidity (HUM), sea-level pressure (PNM), wind direction (WD), and wind speed (WS). Figure \ref{fig:Fig2} shows the raw temperature and relative humidity MD, while the main descriptive statistics are presented in Table \ref{tbl:MVStatics} of the meteorological station located at Córdoba city airport (“SACO” in Figure \ref{fig:Fig1}b) as representative. Additional MD, including all-sky surface shortwave downward irradiance (IRR1), clear-sky surface shortwave downward irradiance (IRR2), all-sky surface shortwave diffuse irradiance (IRR3), and precipitation (PRE) from the CERES SYN1deg satellite product, were obtained from NASA POWER \cite{NASA_POWER2024} for the same location and time period.

\begin{table}[htb!]
  \centering
  \caption{ Main statistics of the meteorological variables recorded by the SMN (SACO) and NASA POWER, all unprocessed. Outliers and the lower and upper bounds were calculated using the three-sigma rule (99.73\%). Units are: Temperature (TEMP) in $^\circ$C, Humidity (HUM) in \%, Pressure (PNM) in hPa, Wind Direction (WD) in $^\circ$, Wind Speed (WS) in km/h, Irradiance (IRR1–IRR3) in MJ/h, and Precipitation (PRE) in mm/h. $\bar{X}$: Mean; SD: Standard Deviation; $\tilde{X}$: Median; IQ: Interquartile Range $Q_3 - Q_1$; $\gamma_1$: Skewness; $\gamma_2$: Kurtosis; $LB$: Lower Bound; $UB$: Upper Bound; $Out$: Outliers; $Mis$: Missing. 
  }  
  \label{tbl:MVStatics}
  \begin{tabular}{|c|c|c|c|c|c|c|c|c|c|c|}
  \hline
  MD &  $\bar{X}$ & SD  & $\tilde{X}$  & IQ  & $\gamma$  &  $\gamma_2$  & $LB$  & $UB$  & $Out$  &   $Mis$\\
  \hline
TEMP & 17.44 & 7.81 & 17.6 & 10.5 & -0.07 & -0.32 & -5.98 & 40.86 & 7 & 245 \\
HUM  & 63.57 & 23.42 & 66.0 & 41.0 & -0.26 & -1.05 & 0.00 & 133.83 & 0 & 252 \\
PNM  & 1014.8 & 6.80 & 1014.4 & 8.9 & 0.23 & 0.08 & 994.4 & 1035.3 & 155 & 245\\
WD   & 140.6 & 116.80 & 140.0 & 210.0 & 0.34 & -1.30 & 0.00 & 360.0 & 0 & 247\\
WS   & 13.0 & 7.90 & 11.0 & 10.0 & 1.10 & 1.33 & 0.00 & 36.7 & 838 & 247 \\
IRR1 & 0.8 & 1.07 & 0.03 & 1.40 & 1.25 & 0.34 & -2.44 & 3.96 & 64 & 0 \\
IRR2 & 1.0 & 1.24 & 0.04 & 1.94 & 0.96 & -0.52 & -2.77 & 4.66 & 0 & 0 \\
IRR3 & 0.2 & 0.33 & 0.03 & 0.38 & 1.59 & 2.17 & -0.76 & 1.23 & 958 & 0\\
PRE  & 1.9 & 13.23 & 0.00 & 0.01 & 26.50 & 1445.26 & -37.83 & 41.54 & 633 & 0\\
\hline
  \end{tabular}
\end{table}

As shown in Table \ref{tbl:MVStatics} (other stations display similar patterns), the SMN records contain missing data. These were treated according to:
\begin{itemize}
    \item Absences of 1–4 hours were imputed by linear interpolation between the immediately preceding and following measurements.
    \item Absences of 5–24 hours were imputed using the average of the same time slot on the previous and following days.
\end{itemize}

Outliers were identified using standard $\sigma$-based approach. For each MD, daily means and standard deviations were computed, and hourly values exceeding $\pm 3\sigma$ from the mean were flagged. Horizontal dotted lines in Figure \ref{fig:Fig2} indicate these thresholds. Outliers in WS were corrected due to obvious typographical errors, whereas outliers in other variables were retained to capture extreme meteorological events, such as heat or cold waves, which can significantly affect electricity demand. Including these extremes in the modeling was essential to evaluate the models’ performance under such conditions.

WD and WS were transformed into cartesian components (zonal (u) and meridional (v) components), converting the circular variable into a vector form. This facilitates physical interpretation of wind patterns and improves machine learning performance by removing the 0°/360° discontinuity. Directions were converted to radians for proper trigonometric calculations, following standard meteorological practice.

\begin{figure}[htb!]
     \centering
     \includegraphics[width=1.0\columnwidth, keepaspectratio=true]{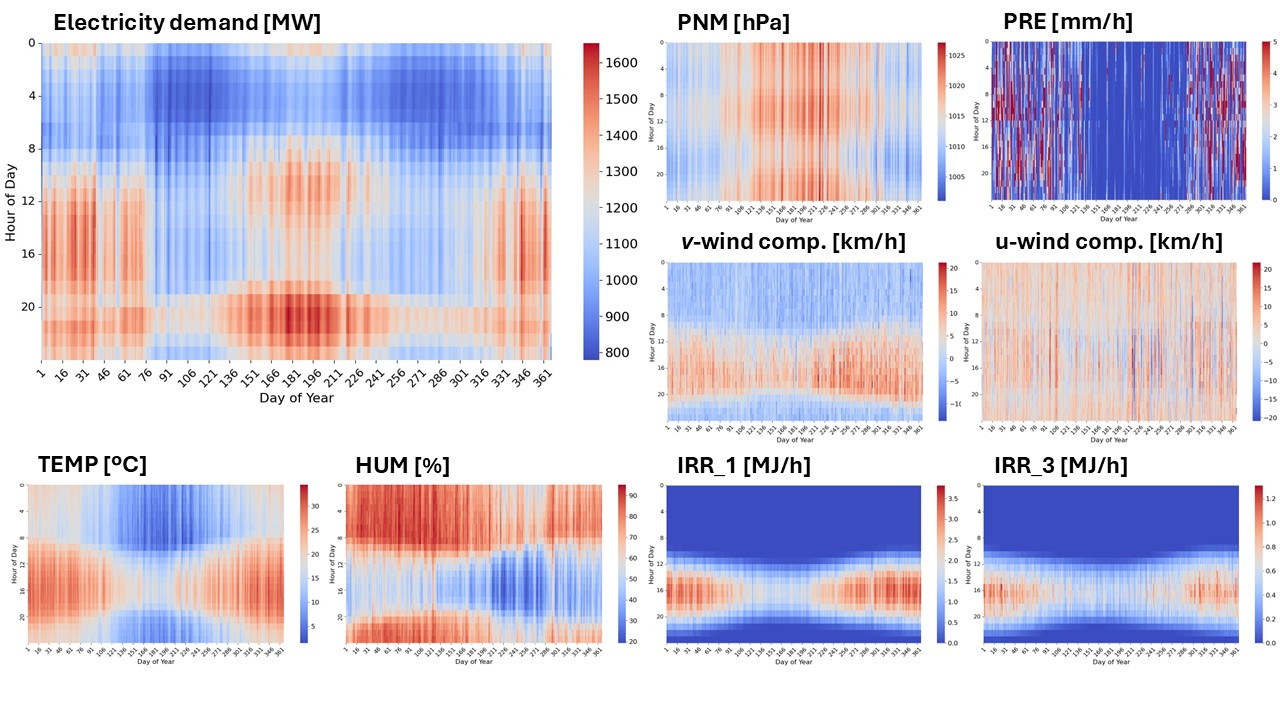}
     \caption{ Heatmaps of electricity demand and meteorological variables averaged hour by hour for each day of the year, using data recorded from 2018 to 2024. The combined evolution reveals cyclic patterns characteristic of the climate in the province of Córdoba, from SMN (SACO) and NASA POWER. In the precipitation panel (PRE), the range has been narrowed to 0–5 mm/h to highlight variability, since precipitation values can reach up to 160 mm/h in short periods.
     }
\label{fig:Fig3}
\end{figure}

MD exhibits strong hourly and seasonal dependencies. Figure \ref{fig:Fig3} presents heatmaps of hourly daily averages from 2018 to 2024 (SACO-SMN), highlighting well-defined cyclic patterns. Temperature shows a pronounced diurnal cycle, with afternoon maxima and early-morning minima, and an amplitude that varies seasonally: sustained high values in summer, lower levels and reduced daily range in winter, and smooth transitions during spring and autumn. Relative humidity displays an inverse pattern, with nighttime maximum and afternoon minimum, and higher averages in colder months. Identifying these patterns (along with the other MD) is crucial for electricity demand forecasting, since extreme temperatures and humidity levels often coincide with peaks in heating or cooling demand.

\subsection{Electric Demand}
Electricity demand data were obtained from the CAMMESA. CAMMESA is an Argentine company composed of both private and public representatives. Its main functions include coordinating electricity dispatch operations, setting wholesale prices, and managing economic transactions through the Argentine Interconnection System (called SADI in Argentine). CAMMESA publishes monthly and annual reports on electricity performance and provides open-access databases \cite{CAMMESA2025}.

\begin{figure}[htb!]
    \centering
    \begin{subfigure}[b]{0.75\textwidth}
        \centering
        \includegraphics[width=\textwidth]{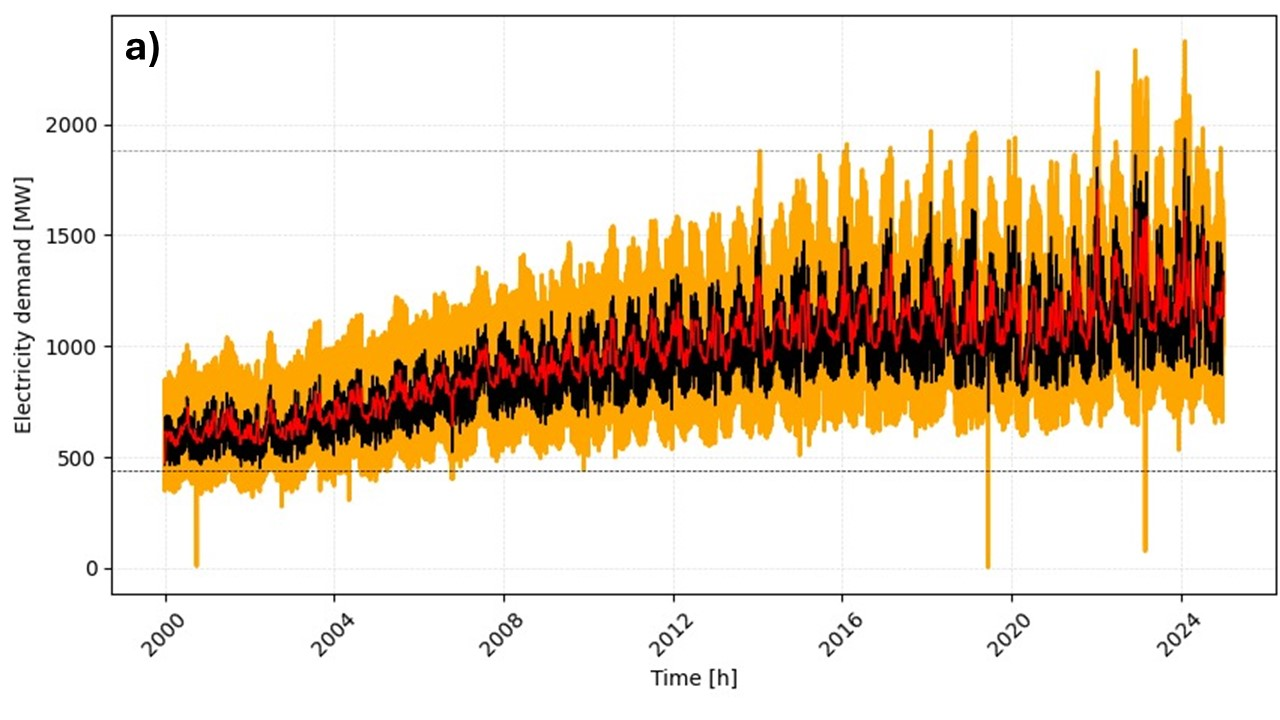}
    \end{subfigure}

    \vspace{0.5em} 

    \begin{subfigure}[b]{0.75\textwidth}
        \centering
        \includegraphics[width=\textwidth]{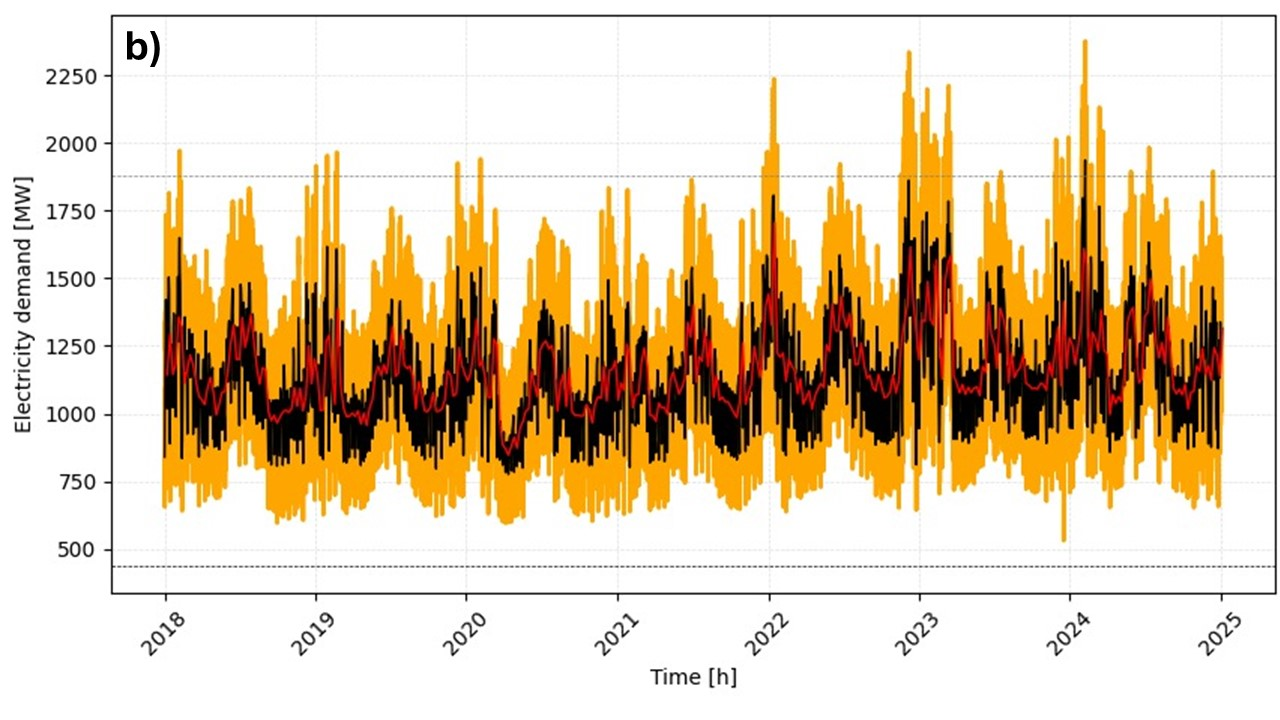}
    \end{subfigure}

    \caption{ Total electricity demand profiles [MW] in Córdoba Province, Argentina. Hourly profile, daily mean, and weekly mean, for a) 01/01/2000–31/12/2024 and b) 01/01/2018–31/12/2024. Dotted lines mark the $3*\sigma$ interval used to identify outliers. In panel b, lower outlier values have been corrected.}
    \label{fig:Fig4}
\end{figure}

Figure \ref{fig:Fig4}a shows the evolution of total electricity demand in the province of Córdoba over the last 25 years, highlighting increases in both the mean value and fluctuations. Figure \ref{fig:Fig4}b focuses on the period 2018–2024, during which demand appears stationary. In this period, the mean demand was 1,157.4 MW, with a standard deviation of 240.8 MW. No missing or duplicated data were identified. Outliers were identified, with limits for this region indicated in Figure \ref{fig:Fig4}b as dotted lines: 434 MW and 1,880 MW. The lower outliers on 16/06/2019 (07:00–16:00) correspond to the largest blackout in Argentina’s history, which also affected Córdoba. The lower outliers on 01/03/2023 (17:00–19:00) were caused by a nationwide high-voltage system failure. Upper outliers correspond to provincial demand records, driven firstly by extreme MD. Lower outliers were rectified, while upper outliers were retained in the model to analyse their relationship with extreme climatic conditions. Figure \ref{fig:Fig3} show the heatmaps of electric demand averaged hour by hour for each day of the year, using data recorded from 2018 to 2024. Note the symmetry of the heatmap around day 185.

The statistical analysis of electricity demand in Córdoba (2018–2024) reveals a bimodal distribution, Section A3 in Supporting Information (SI), attributable to two clearly differentiated consumption regimes, which can be adequately fitted using two Gaussian functions. The Fourier transform, Section A4 in SI, of the demand highlights periodic components associated with daily, weekly, and seasonal cycles, emphasizing the importance of incorporating these frequencies into predictive models (temporal features). Autocorrelation and partial autocorrelation functions further support the existence of a repetitive temporal structure, with peaks every 24 and 168 hours, Section  A5 in SI. Moreover, seasonal decomposition, along with ADF and KPSS tests, Section A6 in SI, indicates that the processed series exhibits stationary behavior over this period, supporting its suitability for modeling.

\begin{figure}[htb!]
     \centering
     \includegraphics[width=1.0\columnwidth, keepaspectratio=true]{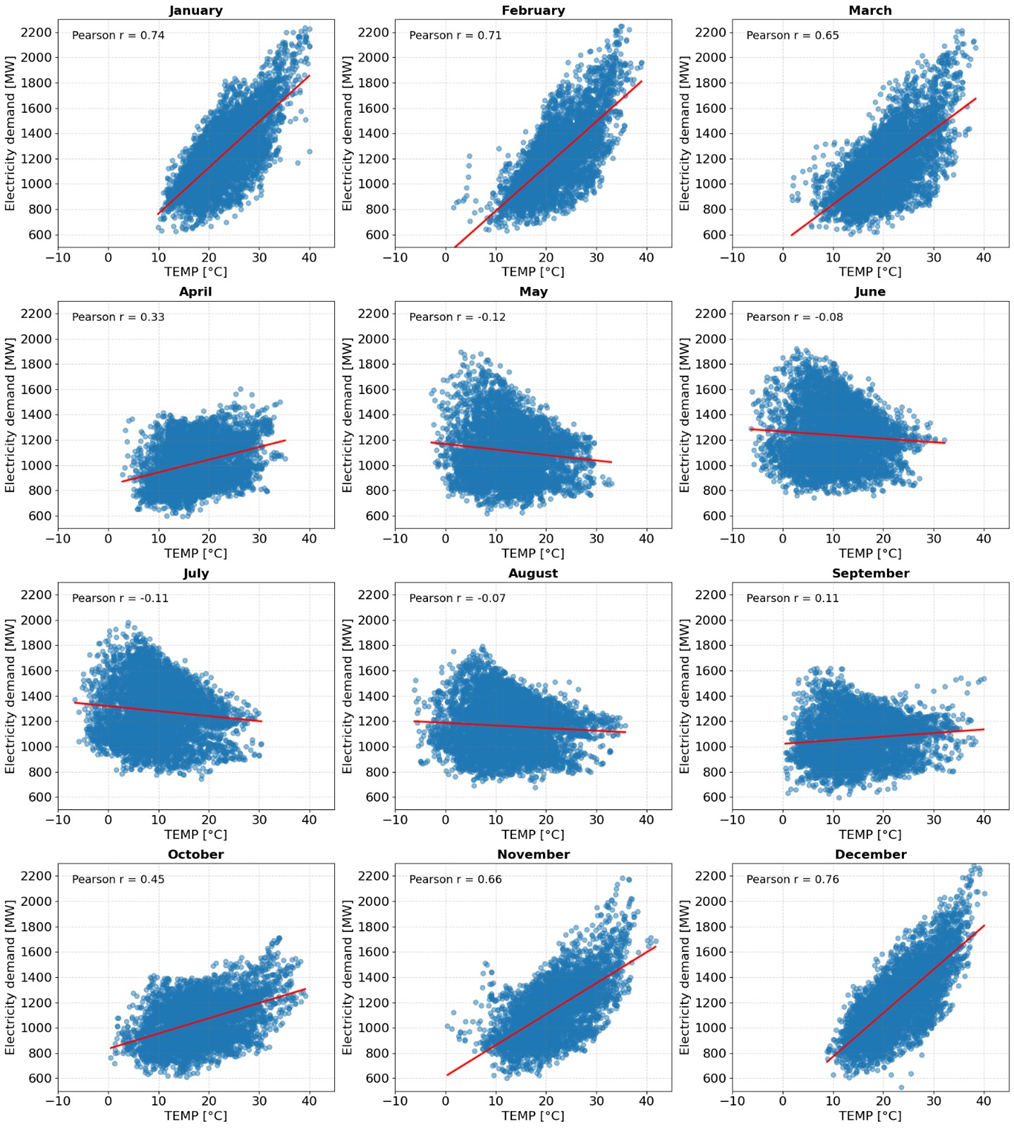}
     \caption{ Electricity demand plotted against temperature for different months of the year, using consistent axis scales to facilitate comparison. The red line indicates the linear fit, with the corresponding Pearson correlation coefficient displayed.
     }
\label{fig:Fig5}
\end{figure}

Figure \ref{fig:Fig5} illustrates the relationship between electricity demand and temperature (from SACO-SMN), disaggregated monthly, revealing an asymmetric response between low- and high-temperature months. This view highlights seasonal variations in the thermal sensitivity of demand, with steeper positive slopes in summer and winter, reflecting the intensive use of heating and cooling systems. For each monthly dataset, a linear fit (red dashed line) and the corresponding Pearson correlation coefficient were computed. Quadratic and cubic fits of temperature were also analyzed, see Section A6 in SI. The divergence between these fits underscores the differing demand responses at low versus high temperatures, with minima of 13.88 $^\circ$C and 15.77 $^\circ$C, respectively. These results suggest that winter heating- and summer cooling-driven consumption follow distinct dynamics, likely influenced by climate-control technologies and consumer behavior. A similar discussion, focusing on the relationship between demand and humidity, is presented in Section A8 of the SI.

\subsection{Periodic Temporal Encodings}
Figure~\ref{fig:Fig2}-\ref{fig:Fig5} highlights the presence of cyclical patterns at daily, weekly, and yearly scales. 
Using raw date and traditional time values can introduce discontinuities, potentially affecting model predictions, for example, the transition from 23:00 to 00:00 when encoding hours as integers, or similar gaps for days of the week or month.

To address this, periodic functions such as sine and cosine were employed, providing a smooth and continuous representation of these cycles. 
Following the FFT analysis (given in Section A4 in SI), sine and cosine encoding were applied at three temporal scales: daily, weekly, and yearly. 
The encoding functions can be written as:

\begin{align}
\text{Daily cycle}\rightarrow  \quad & \text{day\_sin} = \sin\Big( 2 \pi \frac{t}{T_\text{day}} \Big), \quad
\text{day\_cos} = \cos\Big( 2 \pi \frac{t}{T_\text{day}} \Big) \\
\text{Weekly cycle} \rightarrow  \quad & \text{week\_sin} = \sin\Big( 2 \pi \frac{t}{T_\text{week}} \Big), \quad
\text{week\_cos} = \cos\Big( 2 \pi \frac{t}{T_\text{week}} \Big) \\
\text{Yearly cycle} \rightarrow \quad & \text{year\_sin} = \sin\Big( 2 \pi \frac{t}{T_\text{year}} \Big), \quad
\text{year\_cos} = \cos\Big( 2 \pi \frac{t}{T_\text{year}} \Big)
\end{align}

where $t$ represents the elapsed time and $T_i$ with $i = \text{day, week or year}$, denotes the total duration of each respective cycle. A total of six periodic temporal encoding variables were used, allowing the model to capture the inherent cyclical behavior of the demand.

\subsection{Long Short-Term Memory Model}
\begin{figure}[htb!]
     \centering
     \includegraphics[width=1.0\columnwidth, keepaspectratio=true]{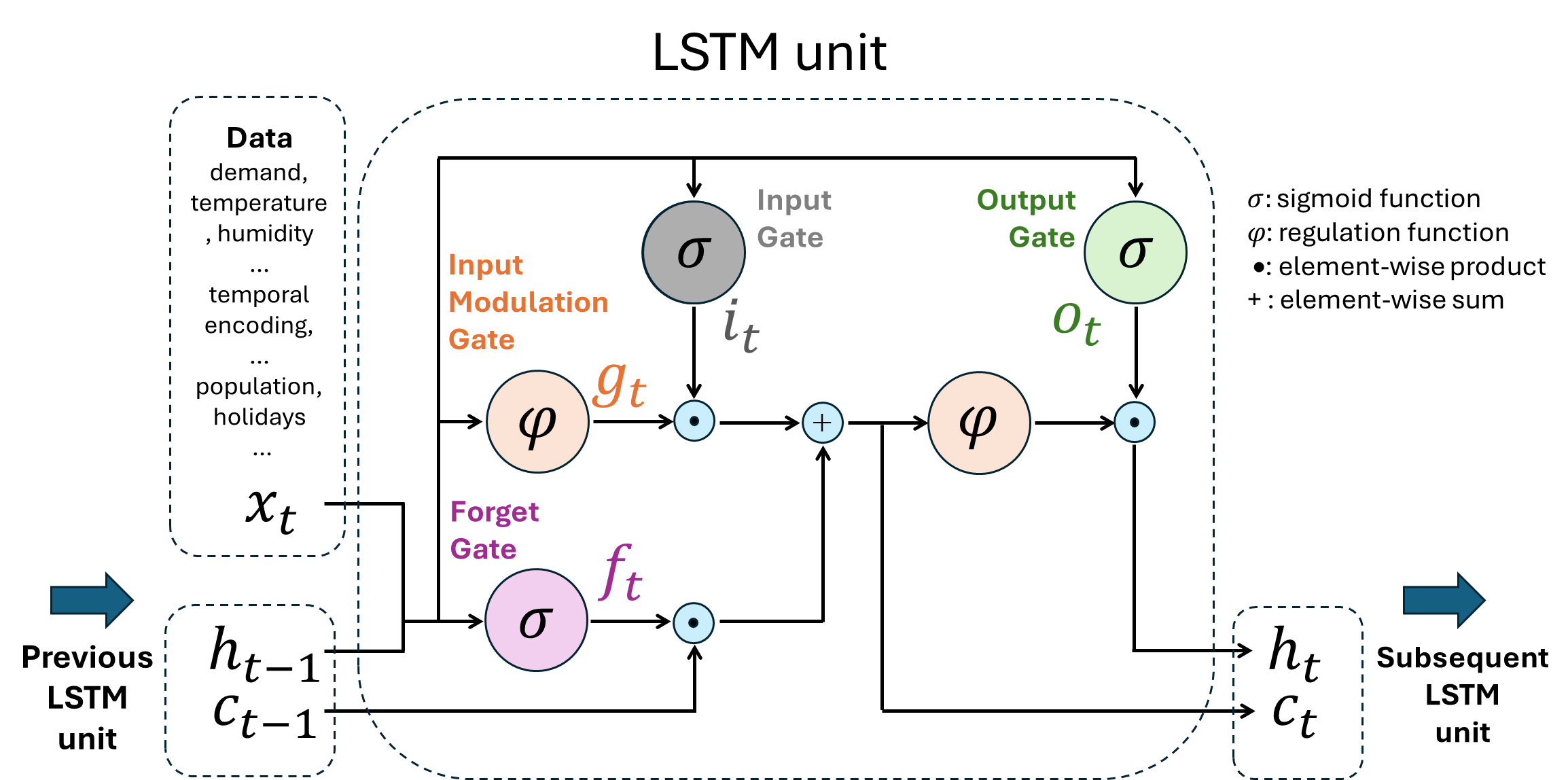}
     \caption{ Internal schematic of an LSTM unit. The cell contains three gates (forget, input, and output) and a modulation function that regulates the update of the internal state $c_t$. The combination of these gates allows the network to maintain and update long-term information, effectively addressing the vanishing gradient problem commonly found in standard recurrent neural networks.
     }
\label{fig:Fig6}
\end{figure}

Hochreiter and Schmidhuber \cite{Hochreiter1997} proposed the Long Short-Term Memory (LSTM) architecture in 1997, which has since been further refined by numerous researchers, becoming one of the most effective structures within recurrent neural networks (RNNs). Its primary objective is to address the vanishing gradient problem, which often affects traditional RNNs when learning long-term dependencies in sequential data. While a basic RNN consists of a series of repetitive modules with simple hidden layers, for instance, a single layer with a sigmoid activation function, LSTM hidden layers feature a more complex internal structure. Each LSTM hidden unit explicitly incorporates a memory cell and a set of gating mechanisms to control information flow. Figure \ref{fig:Fig6} presents a basic schematic of a single LSTM unit, which functions as a repetitive module.

At each time step $t$, the LSTM cell works with the following variables:

\begin{itemize}
  \item $x_t$: the input vector at the current time step $t$.
  \item $h_{t-1}$: the hidden state (output) from the previous time step.
  \item $c_{t-1}$: the memory cell state from the previous time step.
\end{itemize}

These three variables feed into the cell’s internal mechanisms. The outputs are:

\begin{itemize}
  \item $h_t$: the cell output (hidden state) at the current time step $t$.
  \item $c_t$: the updated cell state, passed to the next time step.
\end{itemize}

The \textbf{forget gate} $f_t$ determines which information from the previous cell state $c_{t-1}$ should be discarded. It applies a sigmoid activation:

\begin{equation}
f_t = \sigma\left(W_f \cdot [h_{t-1}, x_t] + b_f \right)
\end{equation}

The value of $f_t \in [0,1]$ acts as a filter controlling how much of the previous memory is retained.

The \textbf{input modulation gate} $g_t$ applies a hyperbolic tangent function to generate candidate content to be stored in the cell:

\begin{equation}
g_t = \tanh\left(W_g \cdot [h_{t-1}, x_t] + b_g \right)
\end{equation}

The \textbf{input gate} $i_t$ decides how much of $g_t$ actually enters the cell:

\begin{equation}
i_t = \sigma\left(W_i \cdot [h_{t-1}, x_t] + b_i \right)
\end{equation}

The \textbf{cell state update} combines the above signals to update the internal state $c_t$:

\begin{equation}
c_t = f_t \cdot c_{t-1} + i_t \cdot g_t
\end{equation}

This allows the cell to retain or replace information in a controlled manner.

The \textbf{output gate} $o_t$ determines which part of the cell state is emitted as the hidden state $h_t$. It is computed as:

\begin{equation}
o_t = \sigma\left(W_o \cdot [h_{t-1}, x_t] + b_o \right)
\end{equation}

Then combined with the updated cell state:

\begin{equation}
h_t = o_t \cdot \tanh(c_t)
\end{equation}

This produces the cell output, which feeds the next step.

This architecture enables the LSTM to store, forget, and update sequential information through gates controlling the data flow. During training, the model learns the coefficients (weights and biases) associated with each gate and internal function. In total, the model trains:

\begin{itemize}
  \item 4 weight matrices: $W_f, W_i, W_g, W_o$
  \item 4 bias vectors: $b_f, b_i, b_g, b_o$
\end{itemize}

LSTM models were constructed, trained, and optimized using the TensorFlow framework (v.2.19.0) and the Keras API (v.3.9.0). Several network architectures were tested to model 24-hour temporal sequences, evaluating a total of ten configurations. Models are specified according to their layer count and the number of neurons per layer (N×N), as summarized in Table \ref{tbl:LSTMArchitectures}. A dense (fully connected) layer is used as the output layer in all architectures. For example, a 64×64 architecture consists of a first (input) layer with 64 neurons, one hidden layer with 64 neurons, and a fully connected dense output layer with a single neuron (not included in the nomenclature).
The ratio of the number of samples to trainable parameters was considered as an indicative measure of overfitting risk. Higher ratios suggest sufficient data for training, whereas ratios closer to one indicate a higher potential for overfitting. The specific values for each architecture are also reported in Table \ref{tbl:LSTMArchitectures} .

\begin{table}[htb!]
\centering
\caption{ Ten LSTM architectures evaluated in this work. A training subset of 49,056 samples was used to assess overfitting risk. All architectures include a fully connected dense layer as the output layer.}
\label{tbl:LSTMArchitectures}
\begin{tabular}{|c|c|c|c|c|}
\hline
\textbf{Model} & \textbf{Layer} & \textbf{Architecture of LSTM} & \textbf{Trainable Parameters} & \textbf{Overfitting Risk} \\
\hline
1  & 2 & 32$\times$32                 & 14,369   & 3.4 \\
2  &   & 64$\times$64                 & 53,313   & 0.9 \\
3  &   & 128$\times$128               & 204,929  & 0.2 \\
\midrule
4  & 3 & 32$\times$32$\times$32       & 22,689   & 2.2 \\
5  &   & 64$\times$64$\times$64       & 86,337   & 0.6 \\
6  &   & 128$\times$128$\times$128    & 336,513  & 0.1 \\
\midrule
7  & 4 & 32$\times$32$\times$32$\times$32   & 31,009   & 1.6 \\
8  &   & 64$\times$64$\times$64$\times$64   & 119,361  & 0.4 \\
9  &   & 128$\times$128$\times$128$\times$128 & 468,097  & 0.1 \\
10 &   & 64$\times$128$\times$128$\times$64 & 300,097  & 0.2 \\
\hline
\end{tabular}
\end{table}

The dataset was divided sequentially into training (80\%), validation (10\%), and test (10\%) sets to preserve the temporal structure. The cleaned dataset comprises 61,368 hourly observations of electricity demand and 18 exogenous features: five meteorological data from SMN (TEMP, HUM, PNM, v-wind and u-wind components), four meteorological data of NASA POWER (IRR1, IRR2, IRR3, PRE), two statistical variables (populations and a binary indicator for workday vs. non-workday), six temporal features (sine and cosine encodings for day, week, and year) and including the demand from the past 24 hours as a feature. The data span from 2018-01-01 to 2024-12-31. That exogenous features were normalized to the [0–1] range.
Each set was processed using 24-hour sliding windows, reflecting the dominant frequency identified via FFT, producing three-dimensional structures suitable for TensorFlow input.

All training and evaluations were performed on a Linux workstation with an $x86$ 64-bit processor (8 physical cores, 16 logical cores, base frequency 4.0 GHz, turbo up to 4.67 GHz) and 16 GB of RAM. Code was developed in Python 3.12.3 (as of February 4, 2025) using virtual environments to ensure reproducibility. Development and execution were carried out in the Windsurf Editor IDE. Runtimes are reported solely to guide model selection in terms of computational efficiency, not for in-depth statistical performance evaluation.

Model performance was primarily evaluated using Root Mean Square Error (RMSE), Mean Absolute Error (MAE), Mean Absolute Percentage Error (MAPE), and the Coefficient of Determination (R²). Additionally, Mean Squared Error (MSE), Weighted Absolute Percentage Error (WAPE), and Mean Absolute Scaled Error (MASE) are also reported: 

\begin{align}
\text{RMSE} &= \sqrt{\frac{1}{n} \sum_{t=1}^{n} (y_t - \hat{y}_t)^2} \\
\text{MAE}  &= \frac{1}{n} \sum_{t=1}^{n} \lvert y_t - \hat{y}_t \rvert \\
\text{MAPE} &= \frac{100}{n} \sum_{t=1}^{n} \left| \frac{y_t - \hat{y}_t}{y_t} \right| \\
R^2 &= 1 - \frac{\sum_{t=1}^{n} (y_t - \hat{y}_t)^2}{\sum_{t=1}^{n} (y_t - \bar{y})^2} \\
\text{MSE}  &= \frac{1}{n} \sum_{t=1}^{n} (y_t - \hat{y}_t)^2 \\
\text{WAPE} &= \frac{\sum_{t=1}^{n} \lvert y_t - \hat{y}_t \rvert}{\sum_{t=1}^{n} y_t} \times 100 \\
\text{MASE} &= \frac{\text{MAE}}{\frac{1}{n-1} \sum_{t=2}^{n} \lvert y_t - y_{t-1} \rvert}
\end{align}

where $y_t$ is the real value at time step $t$, $\hat{y}_t$ is the predicted value, $\bar{y}$ is the mean of the observed values, and $n$ is the total number of observations.

\section{Results and Discussion}

\subsection{Model explainability}
It is instructive to analyze the interpretability of exogenous predictors driving electricity demand. For this purpose, a Random Forest (RF) Regressor was employed, since this ensemble technique not only achieves competitive predictive performance but also provides insights into the relative importance of each feature.  

The model was trained with 18 exogenous variables achieving a satisfactory performance of $R^2 = 0.914$, RMSE = 68.6 MW, and MAPE = 3.99\%. While its accuracy does not reach the level of deep learning architectures such as LSTM, Random Forest offers a key advantage: the ability to quantify feature contributions in an interpretable manner.  

Table \ref{tbl:rf_importance} reports the feature importances extracted from the RF together with the Pearson correlation coefficients between each predictor and electricity demand. As expected, the previous demand emerges as the dominant predictor (64.3\%), reflecting the strong temporal dependence of the series. Among exogenous factors, seasonal harmonics $week\_cos$, $week\_sin$ and TEMP show both high importance and meaningful correlations (\(\rho = 0.32\), \(\rho = -0.06\), \(\rho = 0.24\), respectively). HUM (\(\rho = -0.24\)) and daily harmonics (\(\rho = -0.45\) for $day\_sin$ present significant linear correlations, though their RF importances are relatively small. This divergence highlights that a strong linear association does not necessarily translate into relevance in a nonlinear ensemble model.

Meteorological variables (TEMP, HUM, u/v-wind, $IRR_i$, and PRE) exhibit weak-to-moderate correlations, confirming their role as relevant but not dominant drivers of demand. Socioeconomic indicators, such as population "POP" (\(\rho = 0.17\)) and holidays (\(\rho = -0.12\)), provide additional explanatory power at larger time scales, although their contributions remain minor in magnitude.  

In Section A1 in the SI presents the comparison between the real electricity demand and the RF-predictions, along with the RF-hyperparameters used in its training. The RF-model reproduces the general trend of the demand reasonably well, although it exhibits limited capability in capturing abrupt demand peaks. In addition, Section A2 in SI, provides the full Pearson correlation matrix among the explanatory variables considered in the study.

The explainability analysis fulfills two keys purposes. First, it provides quantitative evidence on the relative weight of exogenous variables, bridging statistical correlations and machine learning importance measures. Second, it motivates the adoption of deep learning approaches (such as LSTM), which, while less transparent, are better suited to capture long-range temporal dependencies and nonlinear interactions, complementing the interpretability offered by ensemble methods.  

\begin{table}[htb!]
\centering
\caption{Feature importance ranking based on Random Forest regression. The mean decrease in impurity (MDI) and normalized importance scores are reported for each predictor.
}
\label{tbl:rf_importance}
\begin{tabular}{|l|c|c|}
\hline
\textbf{Exogenous Variables} & \textbf{MDI Score} & \textbf{Pearson $\rho$} \\
\hline
Previous demand  & 0.643 &  0.79 \\
TEMP             & 0.075 &  0.24 \\
HUM              & 0.006 & -0.24 \\
PNM              & 0.008 & -0.07 \\
u-wind           & 0.009 &  0.09 \\
v-wind           & 0.003 &  0.20 \\
day\_sin         & 0.004 & -0.45 \\
day\_cos         & 0.013 & -0.12 \\
week\_sin        & 0.074 & -0.06 \\
week\_cos        & 0.092 &  0.32 \\
year\_sin        & 0.007 &  0.03 \\
year\_cos        & 0.021 &  0.02 \\
POP              & 0.013 &  0.17 \\
Holiday\         & 0.014 & -0.12 \\
IRR1             & 0.006 &  0.29 \\
IRR2             & 0.003 &  0.29 \\
IRR3             & 0.002 &  0.27 \\
PRE              & 0.007 & -0.02 \\
\hline
\end{tabular}
\end{table}

\subsection{LSTM-Architecture selection}
The selection of an appropriate architecture of the LSTM type is a decisive step when developing predictive models in the energy domain, as it directly impacts their reliability, scalability, and practical applicability. In this study, ten LSTM configurations (Table \ref{tbl:LSTMArchitectures}) were trained using a set composed of 49,056 training records and 6,132 validation set. The ratio between the number of available data points and the number of trainable parameters varied across architectures, ranging from 3.4 to 0.1 (see Table \ref{tbl:LSTMArchitectures} , 5th-column). This imbalance revealed a potential risk of overfitting in the more complex models, which could undermine the robustness of the predictions when applied to unseen operational scenarios (test set). To counteract this risk, a dropout regularization of 20\% was incorporated in the input and hidden layers. This approach was designed to limit overfitting while preserving the capacity of the models to learn meaningful patterns from the data.

Table \ref{tbl:LSTMPerformance} presents the performance metrics for all analysed architectures. Training time was found to increase substantially with both network depth and width. For example, increasing depth while keeping the number of neurons per layer constant (e.g., N×N → N×N×N) resulted in 24–57\% longer training times, whereas doubling the number of neurons per layer at fixed depth (e.g., N×N → 2N×2N) led to 30–78\% longer times. These patterns are consistent with the computational complexity of LSTM networks, where additional layers extend the gradient propagation path and larger layers expand matrix operations quadratically. It is important to note that results are based on a single run per architecture and may be affected by random factors such as weight initialization or batch ordering. We will return to this point in later sections. Nevertheless, the findings provide a clear order-of-magnitude estimate of the trade-off between model complexity and training time, which is critical when designing models intended for reliable and scalable applications in the energy domain. It is worth noting, for instance, that the RF-model was trained in 3 seconds (see the discussion in Section A1 of the SI), although, as mentioned, it does not achieve the performance of the LSTM models.

\begin{table}[htb!]
\centering
\caption{Performance results of different LSTM architectures evaluated across the three sets. Metrics include Mean Absolute Error (MAE), Mean Absolute Percentage Error (MAPE), Root Mean Square Error (RMSE), and the Coefficient of Determination (R²), along with total training time (in seconds).}
\label{tbl:LSTMPerformance}
\begin{tabular}{|c|c|c|c|c|c|c|c|c|c|}
\toprule
Model & Time [s] & Subset & MSE & MAE & RMSE & $R^2$ & MAPE & WAPE & MASE \\
\midrule
\multirow{3}{*}{1} & \multirow{3}{*}{463} & test  & 7749.30  & 68.78 & 88.03 & 0.86 & 5.98 & 9.43 & 0.63 \\
                   &                       & val   & 6550.32  & 57.50 & 80.93 & 0.90 & 4.90 & 9.24 & 0.45 \\
                   &                       & train & 1987.83  & 32.20 & 44.59 & 0.97 & 2.80 & 5.10 & 0.30 \\
\midrule
\multirow{3}{*}{2} & \multirow{3}{*}{732} & test  & 4402.70  & 47.04 & 66.35 & 0.92 & 3.99 & 7.65 & 0.43 \\
                   &                       & val   & 5022.96  & 48.89 & 70.87 & 0.92 & 4.08 & 8.15 & 0.38 \\
                   &                       & train & 1199.66  & 24.97 & 34.64 & 0.98 & 2.17 & 3.99 & 0.23 \\
\midrule
\multirow{3}{*}{3} & \multirow{3}{*}{1105}& test  & 5182.26  & 55.83 & 71.99 & 0.91 & 4.92 & 7.84 & 0.51 \\
                   &                       & val   & 5213.15  & 51.07 & 72.20 & 0.92 & 4.34 & 8.22 & 0.40 \\
                   &                       & train & 1912.78  & 31.67 & 43.74 & 0.97 & 2.69 & 4.88 & 0.30 \\
\midrule
\multirow{3}{*}{4} & \multirow{3}{*}{6789}& test  & 6526.39  & 64.39 & 80.79 & 0.88 & 5.66 & 8.56 & 0.59 \\
                   &                       & val   & 5262.29  & 50.91 & 72.54 & 0.92 & 4.35 & 8.38 & 0.40 \\
                   &                       & train & 2132.47  & 33.59 & 46.18 & 0.96 & 2.90 & 5.21 & 0.31 \\
\midrule
\multirow{3}{*}{5} & \multirow{3}{*}{1059}& test  & 5042.40  & 51.13 & 71.01 & 0.91 & 4.38 & 8.11 & 0.47 \\
                   &                       & val   & 5177.36  & 50.31 & 71.95 & 0.92 & 4.22 & 8.18 & 0.39 \\
                   &                       & train & 1273.27  & 25.84 & 35.68 & 0.98 & 2.26 & 4.11 & 0.24 \\
\midrule
\multirow{3}{*}{6} & \multirow{3}{*}{1744}& test  & 4368.98  & 48.04 & 66.10 & 0.92 & 4.15 & 7.52 & 0.44 \\
                   &                       & val   & 5081.59  & 49.66 & 71.29 & 0.92 & 4.16 & 8.10 & 0.39 \\
                   &                       & train & 1828.89  & 30.70 & 42.77 & 0.97 & 2.60 & 4.79 & 0.29 \\
\midrule
\multirow{3}{*}{7} & \multirow{3}{*}{1007}& test  & 4475.66  & 49.16 & 66.90 & 0.92 & 4.32 & 7.66 & 0.45 \\
                   &                       & val   & 5173.08  & 49.39 & 71.92 & 0.92 & 4.16 & 8.35 & 0.38 \\
                   &                       & train & 1524.41  & 28.39 & 39.04 & 0.97 & 2.49 & 4.49 & 0.27 \\
\midrule
\multirow{3}{*}{8} & \multirow{3}{*}{1312}& test  & 4156.46  & 46.51 & 64.47 & 0.92 & 4.02 & 7.41 & 0.43 \\
                   &                       & val   & 5264.46  & 50.02 & 72.56 & 0.92 & 4.16 & 8.32 & 0.39 \\
                   &                       & train & 1507.50  & 28.05 & 38.83 & 0.97 & 2.42 & 4.43 & 0.26 \\
\midrule
\multirow{3}{*}{9} & \multirow{3}{*}{2333}& test  & 4736.14  & 48.40 & 68.82 & 0.91 & 4.11 & 7.96 & 0.44 \\
                   &                       & val   & 5136.15  & 49.73 & 71.67 & 0.92 & 4.13 & 8.12 & 0.39 \\
                   &                       & train & 1499.29  & 27.59 & 38.72 & 0.97 & 2.39 & 4.49 & 0.26 \\
\midrule
\multirow{3}{*}{10}& \multirow{3}{*}{2318}& test  & 3938.94  & 43.81 & 62.76 & 0.93 & 3.73 & 7.33 & 0.40 \\
                   &                       & val   & 4346.96  & 43.44 & 65.93 & 0.93 & 3.59 & 7.82 & 0.34 \\
                   &                       & train & 714.17   & 19.37 & 26.72 & 0.99 & 1.68 & 3.06 & 0.18 \\
\bottomrule
\end{tabular}
\end{table}

Performance metrics consistently follow the trend: test $<$ validation $<$ training, which is expected since these models are directly trained on the training set, optimizing its parameters to minimize error on those data. The test set, although not used during training and validation, remain independent and is employed to assess model performance across different architectures. The similarity of values across all metrics indicates that the models are robust, an effect further reinforced by the 20\% dropout applied in the first and hidden layers, which helped prevent overfitting.

The 64×128×128×64 configuration shows a significant improvement in predictive capability across all metrics on the test subset. This model achieved the highest coefficient of determination (R² = 0.93) along with the lowest MAPE = 3.73\%, while maintaining a reasonable training time of approximately 40 minutes.

\subsection{LSTM-Hyperparameter optimization}
A hyperparameter optimization stage was conducted to enhance the performance of the LSTM model with the previously selected architecture (64×128×128×64). This search focused on three key training components: the activation function, the optimizer, and the batch size. Different activation functions were evaluated in both recurrent and dense layers, including tanh, ReLU, softmax, and sigmoid, chosen due to their frequent use in the literature for time series tasks and deep networks. For the optimization algorithm, Adam, SGD, and RMSprop were compared to assess the model's sensitivity to different weight update strategies.
For each activation-optimizer pair, batch sizes ranging from 12 to 72, in increments of 12, were explored, given their direct influence on convergence and training stability. In total, 72 configurations were evaluated, amounting to a total training time of 52 hours (approx. 72 x 40 min). Techniques such as early stopping, adaptive learning rate reduction, and selection of the best model based on minimal validation loss were employed. Cross-validation using different splits of the training set (k-fold) was also considered to reduce potential dependency on a single data partition. However, no significant improvements were observed compared to the classical split, and thus this strategy was not implemented.

Table \ref{tbl:LSTM_hyperparams} presents the results of the 10 best hyperparameter combinations (out of a total of 72). The combination Optimizer: Adam + Activation: sigmoid + Batch: 60, achieved the lowest MSE (2906.294), RMSE (53.910), and the highest R² (0.947). The concurrent improvement in both scale-dependent (MSE, RMSE, MAE) and scale-free (MAPE, WAPE, MASE) metrics indicates that this configuration achieves accurate predictions across the full range of values, including extremes, while maintaining strong relative accuracy on small and medium targets.

\begin{table}[htb!]
\centering
\caption{ Performance results of the ten best LSTM models (64×128×128×64) with different hyperparameters, evaluated on the test dataset. All models use the Adam optimizer (omitted from the table)}
\label{tbl:LSTM_hyperparams}
\begin{tabular}{|c|c|c|c|c|c|c|c|c|}
\hline
batch &  activation & MSE & MAE & RMSE & $R^2$ & MAPE & WAPE & MASE \\
\hline
60 & sigmoid & 2906.294 & 39.083 & 53.910 & 0.947 & 3.296 & 6.019 & 0.358 \\
36 & sigmoid & 3013.230 & 39.417 & 54.893 & 0.945 & 3.367 & 6.256 & 0.361 \\
72 & sigmoid & 3196.655 & 40.753 & 56.539 & 0.942 & 3.421 & 6.307 & 0.373 \\
48 & sigmoid & 3334.585 & 41.902 & 57.746 & 0.939 & 3.543 & 6.444 & 0.384 \\
48 & tanh & 3348.091 & 41.579 & 57.863 & 0.939 & 3.589 & 6.650 & 0.381 \\
24 & sigmoid & 3393.779 & 42.751 & 58.256 & 0.938 & 3.638 & 6.466 & 0.392 \\
24 & tanh & 3466.407 & 41.017 & 58.876 & 0.937 & 3.507 & 6.925 & 0.376 \\
84 & sigmoid & 3500.194 & 43.335 & 59.162 & 0.936 & 3.661 & 6.544 & 0.397 \\
36 & tanh & 3778.972 & 42.382 & 61.473 & 0.931 & 3.518 & 7.104 & 0.388 \\
60 & relu & 3862.120 & 45.400 & 62.146 & 0.930 & 3.937 & 7.071 & 0.416 \\
\hline
\end{tabular}
\end{table}

\subsection{Effect of the random seed}
The performance of the best hyperparameter LSTM model, in Table \ref{tbl:LSTM_hyperparams} with a 4-layer architecture (64×128×128×64) + optimizer: Adam + activation: sigmoid and batch size of: 36, 60 and 72, were subjected to thirty independent runs each, varying the random seed used for weight initialization, to assess the stability and robustness of the results against the inherent randomness in training deep NNs. Performance metrics were computed on the test set, and averages and standard deviations were reported for each experiment in Table \ref{tbl:LSTM_seeds}.

\begin{table}[htb!]
\centering
\caption{ Performance results of the best LSTM model (64×128×128×64) with optimized hyperparameters (Optimizer: Adam (0.001) and Activation: Sigmoid), evaluated using different random seeds for model initialization in test set. The mean and standard deviation (SD) across the 30 runs are also reported. Total training time in seconds. BEST’ refers to the models achieving the lowest MAPE and highest $R^2$}

\label{tbl:LSTM_seeds}
\begin{tabular}{|c|c|c|c|c|c|c|c|c|c|c|}
\hline
Batch &   & Time & MSE & MAE & RMSE & $R^2$ & MAPE & WAPE & MASE \\
\hline
36 & BEST &  2380 & 3026.274 & 39.620 & 55.012 & 0.945 & 3.413 & 6.303 & 0.363 \\
   & MEAN &  2396 & 4283.937 & 47.031 & 65.222 & 0.922 & 4.029 & 7.433 & 0.431 \\
   & SD   &  149  & 745.465  & 4.614  & 5.576  & 0.014 & 0.413 & 0.623 & 0.042 \\
\midrule
60 & BEST  & 2141 & 2916.054 & 37.919 & 54.001 & 0.947 & 3.195 & 6.221 & 0.347 \\
   & MEAN  & 1834 & 4799.377 & 50.290 & 68.877 & 0.913 & 4.298 & 7.746 & 0.461 \\
   & SD    & 166  & 1046.356 & 7.013  & 7.567  & 0.019 & 0.618 & 0.754 & 0.064 \\
\midrule
72 & BEST  & 1716 & 3034.894 & 38.387 & 55.090 & 0.945 & 3.203 & 6.337 & 0.352 \\
   & MEAN  & 1747 & 4834.182 & 50.101 & 69.011 & 0.912 & 4.265 & 7.774 & 0.459 \\
   & SD    & 161  & 1213.051 & 7.186  & 8.607  & 0.022 & 0.614 & 0.910 & 0.066 \\
\hline
\end{tabular}
\end{table}

The analysis highlights marked differences in performance among the evaluated configurations. The model trained with a batch size of 36, Adam optimizer, and sigmoid activation function delivered reasonable accuracy, but its variability across runs suggests a certain sensitivity to weight initialization. Increasing the batch size to 60, while keeping the optimizer and activation function unchanged, significantly improved performance: this configuration consistently achieved the lowest error values (MSE, MAE, RMSE, MAPE, WAPE, and MASE) and the highest $R^2$, while also reducing dispersion across independent runs. By contrast, the model with a batch size of 72 achieved a 38\% reduction in training time, which is expected since larger batch sizes accelerate gradient updates. However, this computational advantage came at the cost of higher errors (MSE, RMSE, WAPE) and greater variability compared to the 60-batch configuration. In following, the configuration with batch size 60, Adam optimizer, and sigmoid activation provides the most reliable and accurate predictions, striking the best balance between precision, robustness, and computational efficiency.

\subsection{Performance of the optimized LSTM-model}
\begin{figure}[htb!]
     \centering
     \includegraphics[width=1.0\columnwidth, keepaspectratio=true]{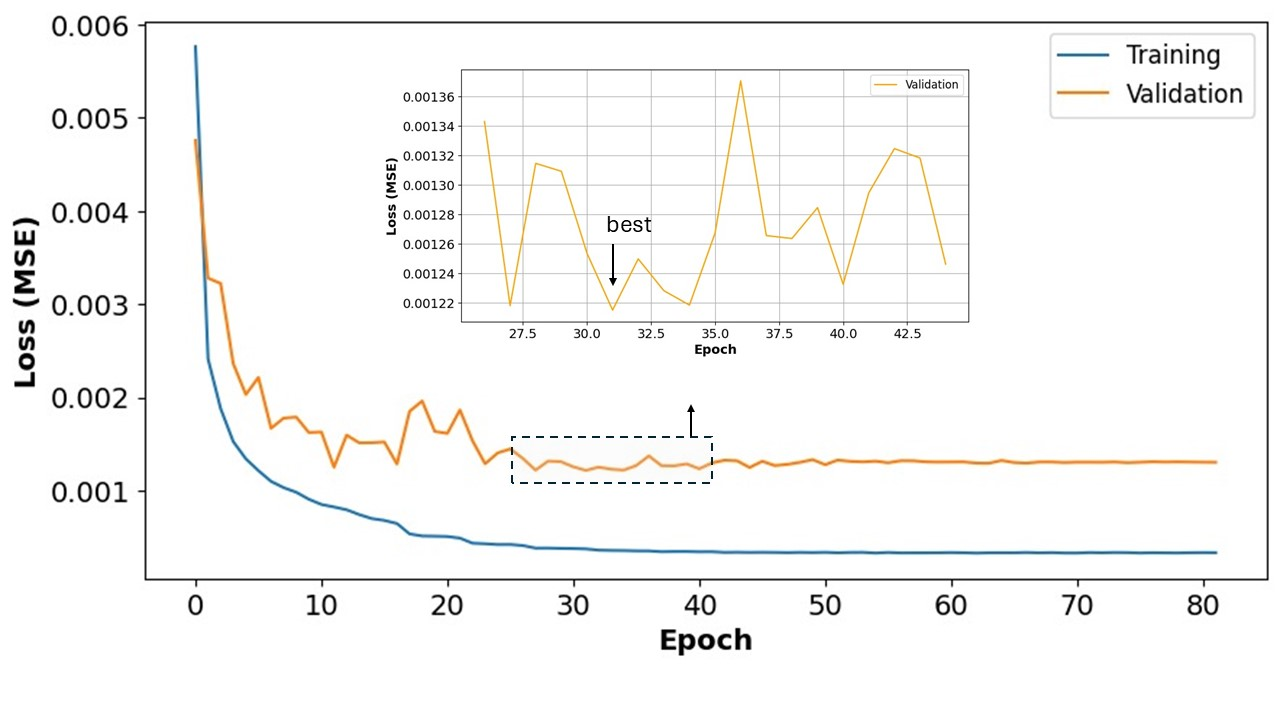}
     \caption{ Evolution of the loss function of the LSTM model over 80 epochs, evaluated on both training and validation sets. The inset shows a magnified view of epochs 25–45. The minimum loss is reached at epoch 31, and the corresponding weights were saved.
     }
\label{fig:Fig7}
\end{figure}

The optimized model is a deep LSTM network with four layers comprising 64×128×128×64 neurons, respectively. Dropout (20\%) was applied to each layer to mitigate overfitting, and a dense layer with a single output neuron was used for regression. The model was trained with a batch size of 60, the Adam optimizer (learning rate = 0.001 with adaptive adjustment), and a sigmoid activation function, using the best random seed in Table \ref{tbl:LSTM_seeds}. The training process is illustrated in Figure \ref{fig:Fig7}. The evolution of the loss function exhibits the typical behavior of neural networks, reaching its best value at epoch 31 (in validation set).

\begin{figure}[htb!]
     \centering
     \includegraphics[width=1.0\columnwidth, keepaspectratio=true]{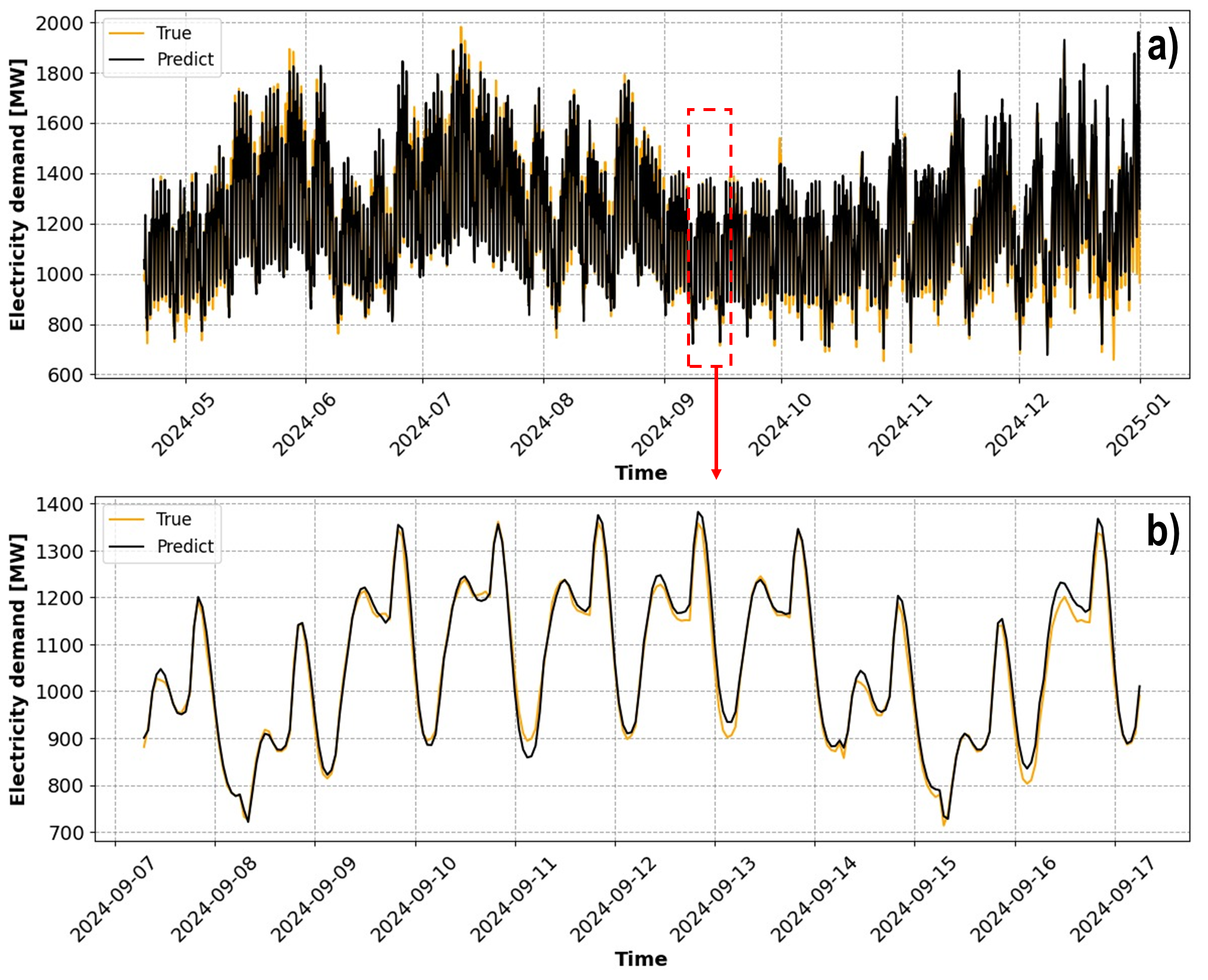}
     \caption{ Comparison between real electricity demand and LSTM model predictions: a) over the entire test set; b) enlarged inset showing model performance over a 10-day time window, highlighting its ability to track daily variations.
     }
\label{fig:Fig8}
\end{figure}

\begin{figure}[htb!]
     \centering
     \includegraphics[width=1.0\columnwidth, keepaspectratio=true]{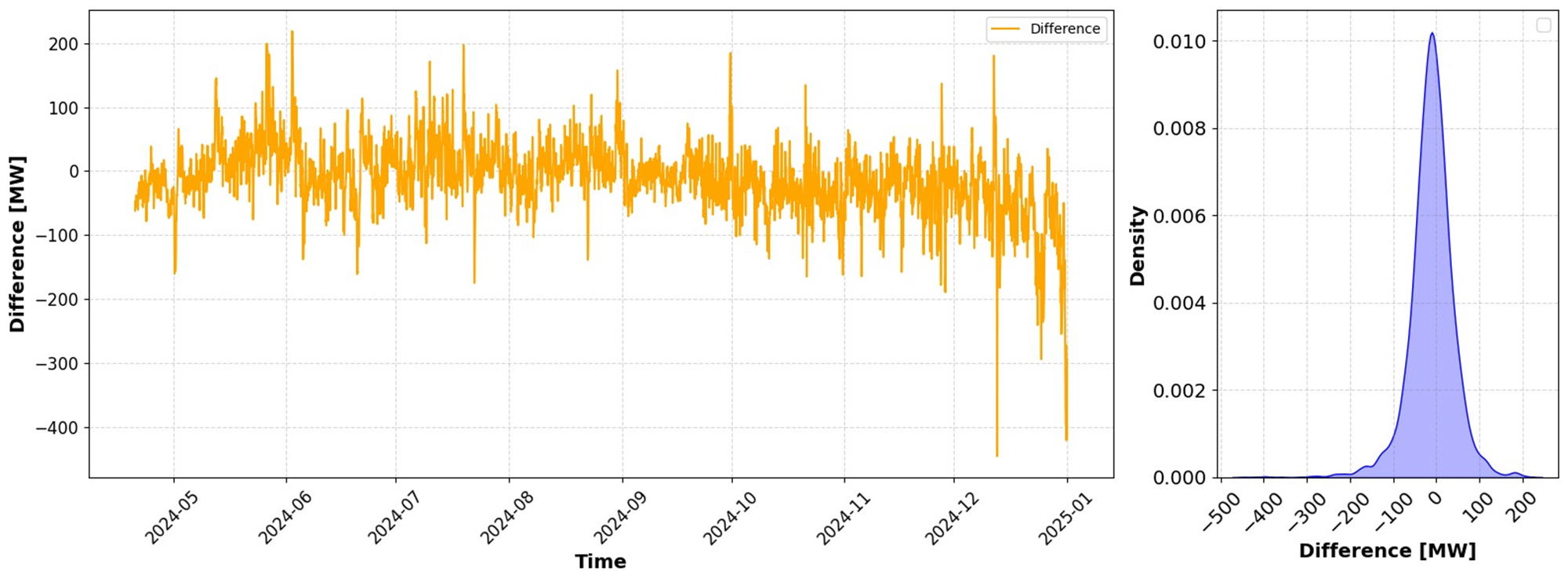}
     \caption{ a) Time series of prediction errors (difference between real and predicted values). b) Probability density function of the prediction errors.
     }
\label{fig:Fig9}
\end{figure}

Figure \ref{fig:Fig8} compares the real electricity demand (True: orange line) with the demand predicted by the LSTM model (Predict: black line) over 6,132 hours (approximately 255 days). The model successfully captures both the trend and seasonality of the hourly series, accurately reproducing short- and long-term variations. The inset provides a zoomed-in view of a specific interval (10 days), showing in greater detail the model’s fit. The curves overlap closely, indicating low deviation between real and predicted values, even in the presence of peaks and troughs characteristic of load behavior. Figure \ref{fig:Fig9} shows the difference between real and predicted electricity demand and the histogram of the distribution of residuals. The distribution is not centered at zero but slightly shifted toward negative values, suggesting that the model tends to slightly overestimate electricity demand (see, e.g., Figure \ref{fig:Fig9}b). Nevertheless, the histogram is symmetric and narrowly concentrated, indicating that errors are generally small and consistent. Table \ref{tbl:LSTM_datasets} reports the standard performance metrics of the LSTM model in all dataset.

\begin{table}[htb!]
\centering
\caption{ LSTM model error metrics for the three datasets (Train, Validation, Test).}
\label{tbl:LSTM_datasets}
\begin{tabular}{lccccccc}
\hline
Dataset & MSE & MAE & RMSE & R² & MAPE & WAPE & MASE \\
\hline
Train      & 776.703  & 19.941 & 27.869 & 0.986 & 1.740 & 3.250 & 0.186 \\
Validation & 4127.660 & 43.478 & 64.247 & 0.936 & 3.569 & 7.389 & 0.339 \\
Test       & 2916.054 & 37.919 & 54.001 & 0.947 & 3.195 & 6.221 & 0.348 \\
\hline
\end{tabular}
\end{table}

Table \ref{tbl:LSTM_datasets} shows that during the training phase, the model achieved performance metrics very close to the test data. On the validation set, the model demonstrated good generalization ability and robustness against overfitting, albeit with a slight increase in error compared to the training set. Finally, on the test data, the model exhibited consistent performance, reaching an $R^2$ of 0.947, similar to the values obtained during training, confirming its stability and predictive capability. The MAPE was 3.195\%, and the MASE was 0.348, both supporting the accuracy and robustness of the model for real-world applications. Overall, these results demonstrate that the proposed LSTM architecture and regularization strategy achieved an optimal balance between fit and generalization, enabling accurate and reliable predictions.

\begin{figure}[htb!]
     \centering
     \includegraphics[width=1.0\columnwidth, keepaspectratio=true]{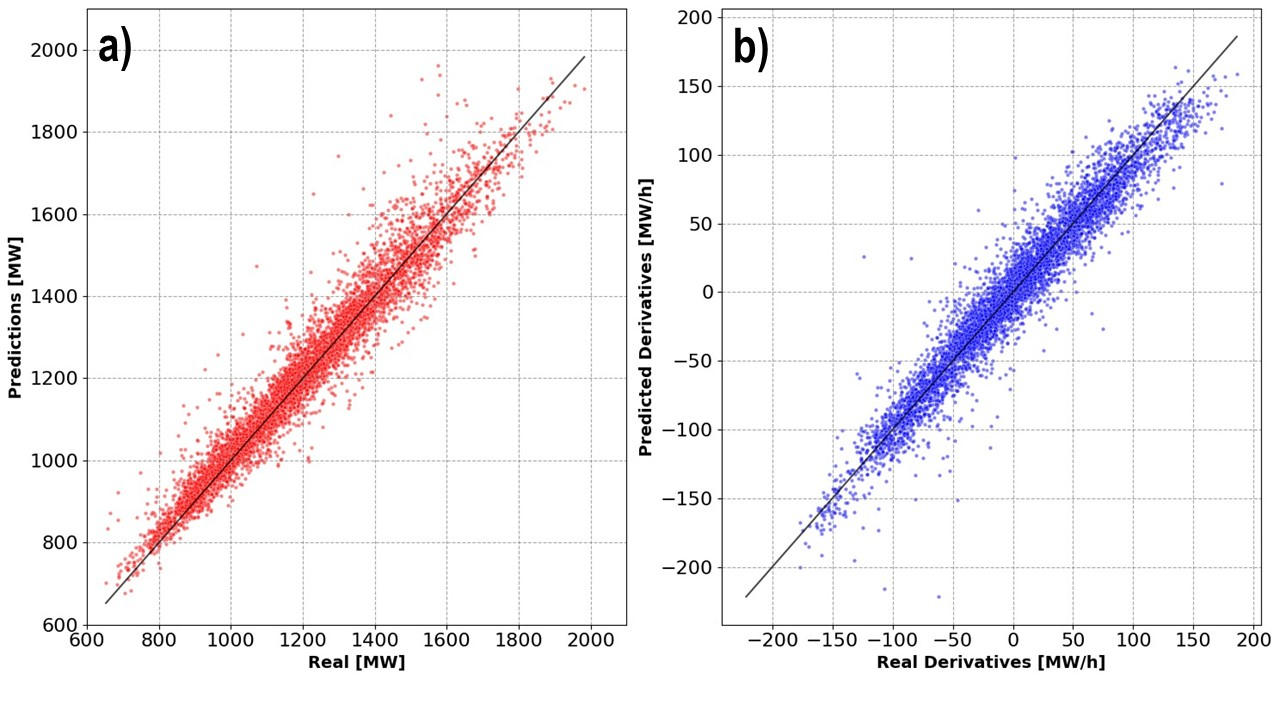}
     \caption{ a) Comparison between real and LSTM model predictions for the electricity demand series, showing high correspondence and estimation accuracy. b) Scatter plot of the discrete temporal derivatives (increments between successive steps) of real and predicted electricity demand, highlighting the model’s ability to capture the dynamic variations of the series. The diagonal y=x (black) are included in each panel to aid visualization.
     }
\label{fig:Fig10}
\end{figure}

Figure \ref{fig:Fig10} presents two scatter plots assessing the performance of the LSTM model in predicting the electricity demand series and its temporal variations. Panel (a) compares real and predicted hourly demand, showing points closely aligned with the $y=x$ diagonal, indicating excellent fit, high predictive accuracy, and no systematic bias. Panel (b) evaluates discrete temporal derivatives (increments between successive steps), where alignment with the diagonal confirms the model’s ability to capture short-term dynamics, despite slightly greater dispersion due to inherent noise in sudden changes. Together, these analyses demonstrate that the LSTM model effectively reproduces both absolute demand levels and their temporal evolution, an essential feature for real-world energy management and operational decision-making.

\subsection{Model variability across temporal contexts}
\begin{figure}[htb!]
     \centering
     \includegraphics[width=1.0\columnwidth, keepaspectratio=true]{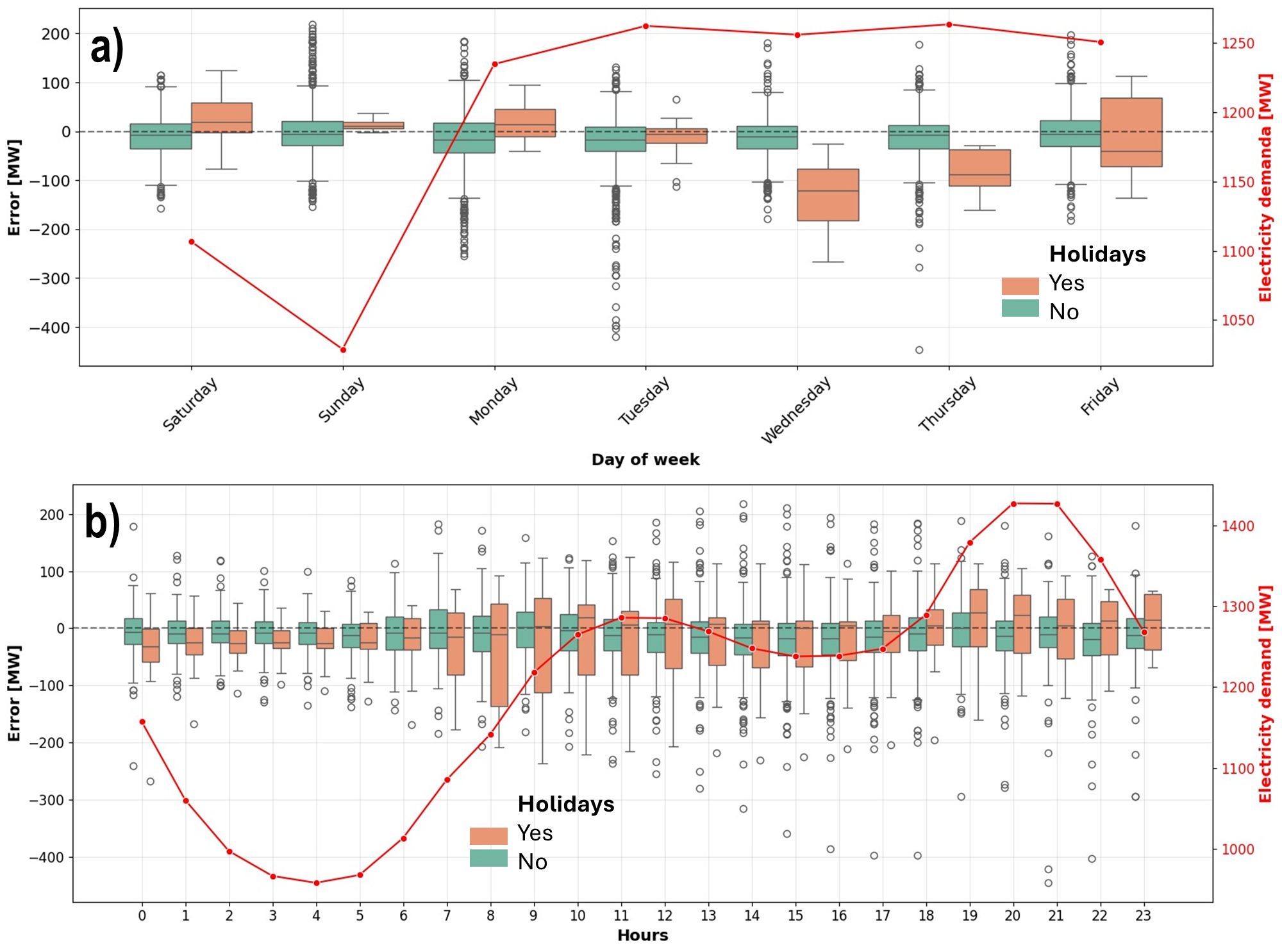}
     \caption{ Prediction error on the test set as a function of: a) day of the week and b) hour of the day, distinguishing between holidays and non-holidays. The secondary (right) axis shows the average demand recorded over the same period, providing context for error variability relative to consumption patterns.
     }
\label{fig:Fig11}
\end{figure}

To evaluate the performance of the LSTM model under different temporal contexts, the error defined as the difference between real ($y_t$) and predicted ($\hat{y}_t$) demand, was analyzed exclusively on the test set. Figure \ref{fig:Fig11}a presents boxplots of the error grouped by day of the week, while Figure \ref{fig:Fig11}b groups errors by hour of the day. In both cases, holidays and special days (called only holidays) and non-holidays are distinguished, and the average demand curve (red, right axis) is included for contextual reference.

In Figure \ref{fig:Fig11}a, holidays (orange bars) exhibit lower error dispersion and almost no outliers, suggesting more homogeneous and predictable consumption patterns due to reduced industrial, commercial, and educational activity. By contrast, working days (green bars) show larger errors, with a negative median (systematic overestimation). Tuesdays stand out with numerous negative outliers, likely linked to higher structural variability in consumption patterns when they coincide with the start of the workweek or when special days are shifted from Monday. A comparable behavior occurs on Thursdays, particularly when the following Friday is classified as a special day.

In Figure \ref{fig:Fig11}b, the hourly analysis reveals a more complex pattern. The model systematically overestimates demand (negative medians), with smaller dispersion on holidays. Outliers remain relatively narrow between 1 and 9 h region, when average demand is low, but their spread increases from 10 to 22 h, peaking during the evening demand maximum (20–21 h).

\subsection{Predictive capability under high and low demand contexts}
Assessing predictive capability under both high- and low-demand contexts is essential, as accuracy during peak periods ensures system reliability, while performance under low-demand conditions reflects sensitivity to baseline variations (for example, in renewable generation planning). Together, these assessments provide a comprehensive measure of model robustness.
For each day $d$ in the test set, the hourly timing errors of daily maxima and minima are defined as:

\begin{align}
\Delta t_{\max, d} &= t_{\max, d}^{\text{real}} - t_{\max, d}^{\text{pred}} \\
\Delta t_{\min, d} &= t_{\min, d}^{\text{real}} - t_{\min, d}^{\text{pred}}
\end{align}

where $t_{\max, d}^{\text{real}}$ and $t_{\min, d}^{\text{real}}$ are the real times of daily maximum and minimum electricity demand respectively, and $t_{\max, d}^{\text{pred}}$ and $t_{\min, d}^{\text{pred}}$ are the predicted times.

\begin{figure}[htb!]
     \centering
     \includegraphics[width=1.0\columnwidth, keepaspectratio=true]{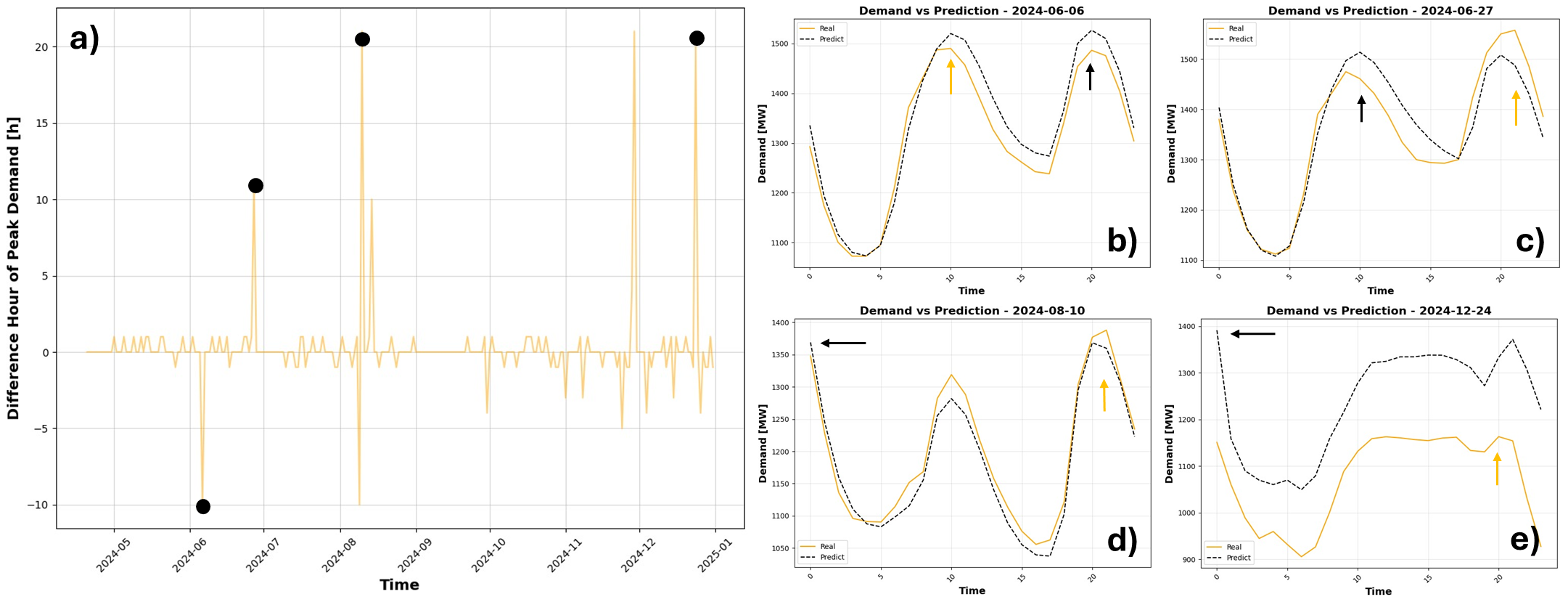}
     \caption{ Hourly difference between the real and predicted time of the daily electricity demand maximum. a) Differences (in hours) over the 255 days of the test set. Panels b), c), d), and e) show examples of the four days with the largest discrepancies, comparing the actual and predicted hourly demand curves. Arrows indicate the time of the maximum value, highlighting significant shifts. Black circles in panel a) correspond to the four days shown in panels b)–e).
     }
\label{fig:Fig12}
\end{figure}

\begin{figure}[htb!]
     \centering
     \includegraphics[width=1.0\columnwidth, keepaspectratio=true]{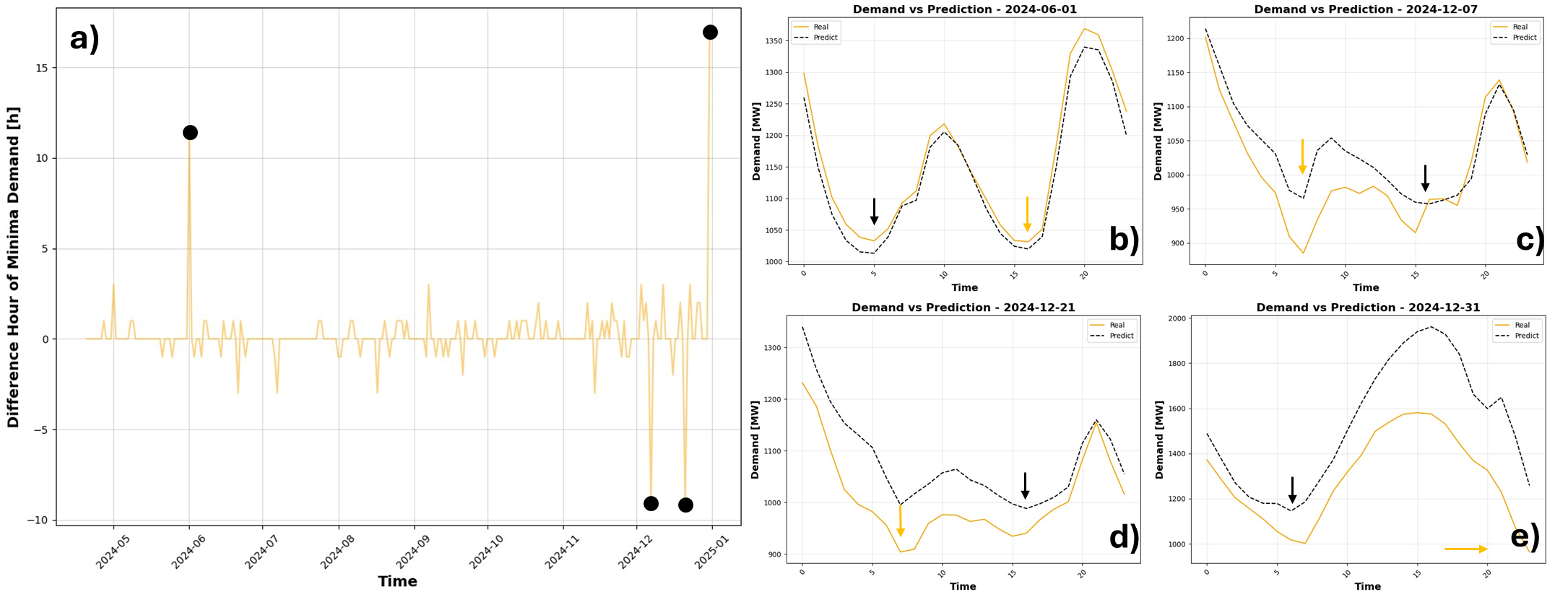}
     \caption{ Hourly difference between the predicted and actual time of the daily electricity demand minimum. a) Differences (in hours) over the 255 days of the test subset. Panels b), c), d), and e) show examples of the four days with the largest discrepancies, comparing the real and predicted hourly demand curves. Arrows indicate the time of the minimum value, highlighting significant shifts. Black circles in panel a) correspond to the four days shown in panels b)–e).
     }
\label{fig:Fig13}
\end{figure}

Results for the 255 days are presented in Figures \ref{fig:Fig12} and \ref{fig:Fig13}. The model predicts the exact hour of the daily maximum in 68.4\% of cases and within ±1 hour in 98.1\% of cases. For daily minima, the exact prediction rate is slightly hight at 69.5\%, but within ±1 hour it lower 91.4\%, likely due to greater variability in nighttime demand patterns. Certain days exhibited significant timing errors, with substantial temporal shifts in the predictions. Real and predicted hourly demands for these cases are shown in Figures \ref{fig:Fig12}b-e and Figure \ref{fig:Fig13}b-e. Two main types of problems were identified: (i) days where maxima or minima are close together, so small prediction errors result in large timing differences (see, for example Figure \ref{fig:Fig12} and \ref{fig:Fig13} b-d) and (ii) days where the demand profile deviates substantially from typical regional patterns (compare with Figure \ref{fig:Fig11}b – red curve), limiting the model’s robustness for accurate prediction (see, for example Figure \ref{fig:Fig12}e and \ref{fig:Fig13}e). For group (i) errors, the model correctly predicts both peaks, but misidentifies local versus global maxima, which can be easily resolved.

\section{Conclusions}
This study demonstrates the ability of LSTM neural network models to efficiently and accurately predict hourly electricity demand, applied to the province of Córdoba, Argentina. By comparing different approaches, it is concluded that the LSTM model, incorporating climatic, temporal, and population variables, achieves the best performance, with a MAPE of 3.195 and a coefficient of determination $R^2$=0.947, results comparable to those reported by other authors for different regions and climates, but using CNN+LSTM hybrid models. Notably, these high-performance values were achieved without modifying the demand data, except for points affected by large-scale power outages impacting the entire province, nor were the exogenous variables altered, except for missing values which were imputed. The inclusion of exogenous variables was crucial to enhance predictive capability, particularly under extreme temperature events that drive high consumption due to heating and cooling. Moreover, the use of periodic temporal encodings enabled proper representation of hourly, weekly, and seasonal cycles, contributing to model stability, as also evidenced by the correspondence between temporal derivatives of actual demand and predictions.

In this work provides a detailed, step-by-step description of the model development and makes both the curated datasets available and full description on hyperparameyters, thus offering a valuable contribution in terms of reproducibility and transparency. The study proposes a robust and scalable tool for regional electricity system management, for which no model in the literature achieves such high accuracy in predicting electricity demand for Argentina or its regions.

Future directions include extending the model to other regions of the country, incorporating probabilistic forecasting approaches, and evaluating the integration of the model into interregional electricity dispatch platforms.

\begin{acknowledgement}
The work was supported by SECyT-UNC [33620190100039CB, 2020] and CONICET [PIP 11220200100725CO, 2021]. Special thanks are extended to Eng. Emiliano Marinozzi from CAMMESA for providing access to historical electricity demand data and for valuable discussions on the electricity market. This work used computational resources from CCAD-UNC, which is part of SNCAD-MinCyT, Argentina.
\end{acknowledgement}

\begin{suppinfo}
Refer to the Supporting information accompanying this paper, which contains detailed information.
\end{suppinfo}

\bibliography{biblio}

\section{Declaration of generative AI and AI-assisted technologies in the manuscript preparation process}
During the preparation of this work the author used ChatGPT (OpenIA, GPT-5) in order to assist with language editing, style refinement, and improvement of clarity in the manuscript. After using this tool, the author reviewed and edited the content as needed and take full responsibility for the content of the published article.

\end{document}